\documentclass{article}%
\usepackage{graphicx}
\usepackage{amsmath}
\usepackage{amsfonts}
\usepackage{amssymb}%
\providecommand{\U}[1]{\protect\rule{.1in}{.1in}}

\begin{document}

\author{Antony Valentini\\Augustus College}

\begin{center}
{\LARGE Statistical anisotropy and cosmological quantum relaxation}

\bigskip

\bigskip

\bigskip\ 

\bigskip\ 

Antony Valentini

\textit{Department of Physics and Astronomy,}

\textit{Clemson University, Kinard Laboratory,}

\textit{Clemson, SC 29634-0978, USA.}

\bigskip

\bigskip

\bigskip
\end{center}

We show that cosmological quantum relaxation predicts an anisotropic
primordial power spectrum with a specific dependence on wavenumber $k$. We
explore some of the consequences for precision measurements of the cosmic
microwave background (CMB). Quantum relaxation is a feature of the de
Broglie-Bohm pilot-wave formulation of quantum theory, which allows the
existence of more general physical states that violate the Born probability
rule. Recent work has shown that relaxation to the Born rule is suppressed for
long-wavelength field modes on expanding space, resulting in a large-scale
power deficit with a characteristic inverse-tangent dependence on $k$. Because
the quantum relaxation dynamics is independent of the direction of the wave
vector for the relaxing field mode, in the limit of weak anisotropy we are
able to derive an expression for the anisotropic power spectrum that is
determined by the power deficit function. As a result, the off-diagonal terms
in the CMB covariance matrix are also determined by the power deficit. We show
that the lowest-order $l-(l+1)$ inter-multipole correlations have a
characteristic scaling with multipole moment $l$. Our derived spectrum also
predicts a residual statistical anisotropy at small scales, with an
approximate consistency relation between the scaling of the $l-(l+1)$
correlations and the scaling of the angular power spectrum at high $l$. We
also predict a relationship between the $l-(l+1)$ correlations at large and
small scales. Cosmological quantum relaxation appears to provide a single
physical mechanism that predicts both a large-scale power deficit and a range
of statistical anisotropies, together with potentially testable relationships
between them.

\bigskip

\bigskip

\bigskip

\bigskip

\bigskip

\bigskip

\bigskip

\bigskip

\bigskip

\bigskip

\bigskip

\bigskip

\bigskip

\bigskip

\bigskip

\bigskip

\bigskip

\bigskip

\bigskip

\bigskip

\bigskip

\bigskip

\bigskip

\bigskip

\bigskip

\section{Introduction}

Tentative evidence has accumulated for the existence of primordial anomalies
as deduced from observations of the cosmic microwave background (CMB), in
particular a large-scale power deficit together with a range of violations of
statistical isotropy (at both large and small scales). An exhaustive analysis
and survey has recently appeared in the \textit{Planck} 2015 data release
\cite{Planck15-XI-PowerSpec,Planck15-XVI-IsoStats,Planck15-XX-Inflation}. As
emphasised by the Planck team, in the absence of an established theoretical
framework that predicts such effects the statistical significance of the
anomalies remains difficult to assess. Are they mere fluctuations or do they
reflect new physics? The absence of established predictive models also makes
it difficult to judge what to look for in such a large and multi-faceted data
set. However, if a single mechanism could be found that explains several of
the anomalies at once as well as predicting relationships between them, one
might obtain a more decisive statistical significance (for or against the model).

Some theoretical models address the large-scale power deficit. This anomaly
might be explained by corrections to the quantum inflationary vacuum state
induced by a radiation-dominated pre-inflationary phase \cite{PK2007,WN08}, by
a suitable period of fast rolling for the inflaton field \cite{CL03,GKMNS15},
or by a form of multifield inflation \cite{SSK14}. Other models address the
statistical anisotropy. This multi-faceted anomaly, or aspects of it, might be
explained in a number of ways, such as a marginally-open universe with a
finite curvature scale (which can induce a dipolar modulation of the
temperature anisotropy) \cite{LC13} or the pre-inflationary decay of a false
vacuum with a smaller number of large dimensions than the current vacuum
\cite{S15}. Statistical anisotropy can also arise from a coupling between the
inflaton and a vector field \cite{BMPS15,S12,NKY14}. Power asymmetry can be
generated by a multifield inflationary (curvaton) model \cite{EKC08} or by a
primordial domain wall \cite{JAFSW14}. Finally, a few models address both
kinds of anomaly. Both a power deficit and a statistical anisotropy can be
generated by a strongly-anisotropic pre-inflationary phase described by a
Kasner geometry \cite{BPM15} or by a direction-dependent inflaton field with
an induced anisotropic Hubble parameter \cite{MS15}. It remains to be seen if
any of these models can provide a fully satisfactory account of the observed
anomalies and if the detailed predictions they make stand up to close
comparison with the data.

In this paper we are concerned with a physical mechanism -- cosmological
quantum relaxation
\cite{AV91a,AV91b,AV92,AV96,AV01,AV07,AV08,AV09,AV10,CV13,CV15} -- which
predicts both a power deficit and statistical anisotropy at large scales, as
well as a residual statistical anisotropy at small scales, and which
furthermore predicts testable relationships between the two kinds of anomaly.

The de Broglie-Bohm pilot-wave formulation of quantum theory
\cite{deB28,BV09,B52a,B52b,Holl93} provides a natural generalisation of the
usual quantum formalism, with probabilities potentially different from those
predicted by the Born rule \cite{AV91a,AV91b,AV92,AV96,AV01,PV06}. On this
view, the Born rule is not a law but only describes a particular state of
statistical equilibrium. Conventional quantum physics is then merely an
effective theory of `quantum equilibrium'. A wider physics of quantum
nonequilibrium -- with violations of the Born rule and associated new
phenomena -- is possible in principle
\cite{AV91a,AV91b,AV92,AV96,AV01,AV02,AV07,AV08,AV08a,AV09,AV10,AVPwtMw,PV06}.
This wider physics may have existed in the early universe
\cite{AV91a,AV91b,AV92,AV96,AV01}, in which case it could leave discernible
traces in the CMB \cite{AV07,AV10,CV13,CV15} and possibly even survive until
today for some relic cosmological particles \cite{AV01,AV07,AV08,UV15}.

In pilot-wave theory, one may consider an expanding universe that begins in a
state of quantum nonequilibrium -- that is, with non-Born rule probabilities
at the earliest times \cite{AV92,AV96,AV01,AV07,AV08,AV10}. If one begins with
such a state, for example for a representative scalar field on expanding
space, the dynamics of pilot-wave field theory allows us to evolve the state
forward in time. Extensive numerical simulations have shown that
short-wavelength field modes evolve rapidly to the Born rule
\cite{VW05,EC06,TRV12,SC12,ACV14}. We would therefore expect local laboratory
physics to obey the Born rule. This expectation is consistent with what is
observed in (for example) atomic physics and high-energy scattering
experiments. However, a combination of theoretical analysis and further
simulations has shown that long-wavelength (super-Hubble) field modes evolve
more slowly and need not reach equilibrium \cite{AV07,AV08,AV10,CV13,CV15}.

If one takes pilot-wave theory seriously on its own terms, one may reasonably
expect the existence of departures from the Born rule at the longest
wavelengths. Cosmologically, there could exist a power deficit in the CMB at
the largest scales induced by an early nonequilibrium suppression of Born-rule
quantum noise. This possibility has been studied in some detail
\cite{AV07,AV08,AV10,CV13,CV15}. We have been unable to predict the current
lengthscale for such effects, since this depends on unknown details of the
early expansion (such as the number of inflationary e-folds). But we are able
to predict the `shape' of the deficit. Recent detailed simulations have shown
that the nonequilibrium deficit will have an approximately inverse-tangent
dependence on wavenumber $k$ \cite{CV15}, a prediction which may be compared
with current data \cite{PVV15}.

In this paper it will be shown that we are also able to predict associated
statistical anisotropies and certain features thereof. By a straightforward
argument, the form of the anisotropic spectrum follows (in the limit of weak
anisotropy) from the inverse-tangent deficit function which has already been
obtained from numerical simulations. We then have a single physical mechanism
-- quantum relaxation on expanding space -- that makes detailed predictions
for both kinds of anomaly. As we shall see, the mechanism predicts that the
statistical anisotropies will be related to the power deficit in quantitative
ways which may in principle be testable. Furthermore, the mechanism can
naturally induce residual statistical anisotropies at small scales.

In Section 2 we outline the essential formalism required for a discussion of
anisotropic primordial power spectra and we provide a brief summary of the
statistical anomalies in the CMB. In Section 3 we first summarise the evidence
that cosmological quantum relaxation predicts a large-scale power deficit that
varies as an inverse-tangent with wavenumber $k$. We then show how to
generalise this result to an anisotropic power spectrum, where in pilot-wave
field theory nonequilibrium anisotropy can exist even in the (conventionally
isotropic) Bunch-Davies vacuum. In the limit of weak anisotropy, we derive an
expression for the anisotropic primordial power spectrum which is determined
by the inverse-tangent power deficit. In Section 4 we express the CMB
covariance matrix (for the harmonic coefficients of the temperature
anisotropy) in terms of our anisotropic primordial power spectrum and we show
how the lowest-order off-diagonal corrections are determined by our
inverse-tangent deficit function. In Section 5 we discuss some anisotropic
signatures of cosmological quantum relaxation. We show that the anisotropic
correlations between neighbouring CMB multipoles, as quantified by the
$l-(l+1)$ terms in the covariance matrix, scale in a particular way with $l$
-- where the scaling depends on the inverse-tangent form of the power deficit
function. Furthermore, we show that our anisotropic power spectrum predicts a
residual statistical anisotropy even at small scales. We also derive an
approximate consistency relation between the scaling of the $l-(l+1)$
covariance matrix elements and the scaling of the angular power spectrum
$C_{l}$ in the limit of high $l$. Finally, we briefly consider how
cosmological quantum relaxation might account for the anomalous alignment
observed between multipoles at very low $l$. In Section 6 we summarise our
results, compare them with related work, and draw our conclusions.

\section{Statistical anomalies in the cosmic microwave background}

In this section we first briefly summarise some essential formalism. Then we
consider the role played by isotropic primordial power spectra and how the
assumption of isotropy may be dropped. Finally, we outline the apparent
statistical anomalies in the CMB as they are currently understood.

\subsection{Primordial fluctuations and the CMB}

A primordial curvature perturbation $\mathcal{R}_{\mathbf{k}}$ (working in
Fourier space) generates coefficients \cite{LR99}%
\begin{equation}
a_{lm}=\frac{i^{l}}{2\pi^{2}}\int d^{3}\mathbf{k}\ \mathcal{T}(k,l)\mathcal{R}%
_{\mathbf{k}}Y_{lm}(\mathbf{\hat{k}}) \label{alm}%
\end{equation}
in the harmonic expansion%
\begin{equation}
\frac{\Delta T(\mathbf{\hat{n}})}{\bar{T}}=\sum_{l=2}^{\infty}\sum_{m=-l}%
^{+l}a_{lm}Y_{lm}(\mathbf{\hat{n}}) \label{har}%
\end{equation}
of the measured CMB temperature anisotropy $\Delta T(\mathbf{\hat{n}})\equiv
T(\mathbf{\hat{n}})-\bar{T}$ (where $\bar{T}$ is the average temperature over
the sky). Here $\mathcal{T}(k,l)$ is the transfer function and the unit vector
$\mathbf{\hat{n}}$ specifies a point on the sky.

It is usually assumed that the function $\Delta T(\mathbf{\hat{n}})$ observed
by us is a single realisation from an underlying theoretical ensemble with an
associated probability distribution. It is common to assume statistical
isotropy for the ensemble. This implies that the two-point correlation
function $\left\langle \Delta T(\mathbf{\hat{n}})\Delta T(\mathbf{\hat{n}%
}^{\prime})\right\rangle $ -- where $\left\langle ...\right\rangle $ denotes
an ensemble average -- depends only on the angle between $\mathbf{\hat{n}}$
and $\mathbf{\hat{n}}^{\prime}$. Using properties of spherical harmonics, this
in turn implies the standard expression \cite{HS05}%
\begin{equation}
\left\langle a_{l^{\prime}m^{\prime}}^{\ast}a_{lm}\right\rangle =\delta
_{ll^{\prime}}\delta_{mm^{\prime}}C_{l} \label{iso}%
\end{equation}
for the covariance matrix $\left\langle a_{l^{\prime}m^{\prime}}^{\ast}%
a_{lm}\right\rangle $ of the harmonic coefficients, where%
\begin{equation}
C_{l}\equiv\left\langle \left\vert a_{lm}\right\vert ^{2}\right\rangle
\end{equation}
is the angular power spectrum.

It is also commonly assumed that the theoretical ensemble for $\mathcal{R}%
_{\mathbf{k}}$ is statistically homogeneous. This implies that the real-space
two-point correlation function $\left\langle \mathcal{R}(\mathbf{x}%
)\mathcal{R}(\mathbf{x}^{\prime})\right\rangle $ depends only on the
separation between $\mathbf{x}$ and $\mathbf{x}^{\prime}$. This in turn
implies that $\left\langle \mathcal{R}_{\mathbf{k}}\mathcal{R}_{\mathbf{k%
\acute{}%
}}\right\rangle e^{-i(\mathbf{k}+\mathbf{k}%
\acute{}%
)\cdot\mathbf{d}}=\left\langle \mathcal{R}_{\mathbf{k}}\mathcal{R}_{\mathbf{k%
\acute{}%
}}\right\rangle $ for arbitrary $\mathbf{d}$, which implies the standard
relation%
\begin{equation}
\left\langle \mathcal{R}_{\mathbf{k%
\acute{}%
}}^{\ast}\mathcal{R}_{\mathbf{k}}\right\rangle =\delta_{\mathbf{kk}%
\acute{}%
}\left\langle \left\vert \mathcal{R}_{\mathbf{k}}\right\vert ^{2}\right\rangle
\ . \label{RkRk}%
\end{equation}
Using (\ref{RkRk}) it follows from (\ref{alm}) that%
\begin{equation}
C_{l}=\frac{1}{2\pi^{2}}\int_{0}^{\infty}\frac{dk}{k}\ \mathcal{T}%
^{2}(k,l)\mathcal{P}_{\mathcal{R}}(k)\ , \label{Cl2}%
\end{equation}
where%
\begin{equation}
\mathcal{P}_{\mathcal{R}}(k)\equiv\frac{4\pi k^{3}}{V}\left\langle \left\vert
\mathcal{R}_{\mathbf{k}}\right\vert ^{2}\right\rangle \label{PPS}%
\end{equation}
is the primordial power spectrum (with $V$ a normalisation volume).

During the inflationary expansion, an inflaton perturbation $\phi_{\mathbf{k}%
}$ generates a curvature perturbation $\mathcal{R}_{\mathbf{k}}\propto
\phi_{\mathbf{k}}$, where $\phi_{\mathbf{k}}$ is evaluated at a time a few
e-folds after the mode exits the Hubble radius \cite{LL00}. Observational
constraints on the $C_{l}$'s imply constraints on $\mathcal{P}_{\mathcal{R}%
}(k)$ and hence (since $\mathcal{R}_{\mathbf{k}}\propto\phi_{\mathbf{k}}$)
constraints on the primordial variance $\left\langle \left\vert \phi
_{\mathbf{k}}\right\vert ^{2}\right\rangle $ for $\phi_{\mathbf{k}}$. Thus
observations of the CMB today imply constraints on quantum fluctuations in the
very early universe \cite{Muk05,PU09}.

In an ideal Bunch-Davies vacuum the inflaton perturbations $\phi_{\mathbf{k}}$
will have (in the limit $k/a<<H$) a quantum-theoretical variance%
\begin{equation}
\left\langle |\phi_{\mathbf{k}}|^{2}\right\rangle _{\mathrm{QT}}=\frac
{V}{2(2\pi)^{3}}\frac{H^{2}}{k^{3}}\ , \label{B-D}%
\end{equation}
where $H$ is the (approximately constant) Hubble parameter. This implies a
scale-free power spectrum%
\begin{equation}
\mathcal{P}_{\phi}^{\mathrm{QT}}(k)\equiv\frac{4\pi k^{3}}{V}\left\langle
\left\vert \phi_{\mathbf{k}}\right\vert ^{2}\right\rangle _{\mathrm{QT}}%
=\frac{H^{2}}{4\pi^{2}}\ .
\end{equation}
The quantity $\left\langle \left\vert \phi_{\mathbf{k}}\right\vert
^{2}\right\rangle _{\mathrm{QT}}$ is calculated from quantum field theory. In
the slow-roll approximation we then obtain an approximately scale-free
quantum-theoretical power spectrum%
\begin{equation}
\mathcal{P}_{\mathcal{R}}^{\mathrm{QT}}(k)\equiv\frac{4\pi k^{3}}%
{V}\left\langle \left\vert \mathcal{R}_{\mathbf{k}}\right\vert ^{2}%
\right\rangle _{\mathrm{QT}}\approx\mathrm{const}. \label{PRQT}%
\end{equation}
for $\mathcal{R}_{\mathbf{k}}$. Of course, $H$ is only approximately constant
during inflation and there will in fact be a mild dependence of $\mathcal{P}%
_{\mathcal{R}}^{\mathrm{QT}}(k)$ on $k$. This simple treatment suffices for
our purposes.

\subsection{Isotropic and anisotropic primordial power spectra}

We have presented the usual justification for the expressions (\ref{iso}) and
(\ref{Cl2}) -- assuming that $\Delta T(\mathbf{\hat{n}})$ is statistically
isotropic and that $\mathcal{R}_{\mathbf{k}}$ is statistically homogeneous.
However, (\ref{iso}) and (\ref{Cl2}) may be derived directly from (\ref{alm})
assuming that $\left\langle \left\vert \mathcal{R}_{\mathbf{k}}\right\vert
^{2}\right\rangle $ is independent of the direction of the wave vector
$\mathbf{k}$ (and that $\mathcal{R}_{\mathbf{k}}$ is statistically
homogeneous). This provides a convenient starting point for generalisation,
with $\left\langle \left\vert \mathcal{R}_{\mathbf{k}}\right\vert
^{2}\right\rangle $ dependent on the direction of $\mathbf{k}$.

If $\mathcal{R}_{\mathbf{k}}$ is statistically homogeneous we have the
relations (\ref{RkRk}) or%
\[
\left\langle \mathcal{R}_{\mathbf{k%
\acute{}%
}}^{\ast}\mathcal{R}_{\mathbf{k}}\right\rangle =\frac{(2\pi)^{3}}{V}\delta
^{3}(\mathbf{k}-\mathbf{k}^{\prime})\left\langle \left\vert \mathcal{R}%
_{\mathbf{k}}\right\vert ^{2}\right\rangle
\]
(using $V\delta_{\mathbf{kk}%
\acute{}%
}=(2\pi)^{3}\delta^{3}(\mathbf{k}-\mathbf{k}^{\prime})$ for $V\rightarrow
\infty$). From (\ref{alm}) we then have%
\begin{align}
\left\langle a_{l^{\prime}m^{\prime}}^{\ast}a_{lm}\right\rangle  & =\frac
{1}{2\pi^{2}}(-i)^{l^{\prime}}i^{l}\int_{0}^{\infty}\frac{dk}{k}%
\ \mathcal{T}(k,l^{\prime})\mathcal{T}(k,l)(4\pi k^{3}/V)\label{covmat1}\\
& \times\int d\Omega\ \left\langle \left\vert \mathcal{R}_{\mathbf{k}%
}\right\vert ^{2}\right\rangle Y_{l^{\prime}m^{\prime}}^{\ast}(\mathbf{\hat
{k}})Y_{lm}(\mathbf{\hat{k}})\nonumber
\end{align}
(where $d\Omega$ is a solid-angle element in $\mathbf{k}$-space).

If we assume that $\left\langle \left\vert \mathcal{R}_{\mathbf{k}}\right\vert
^{2}\right\rangle $ is independent of the direction of $\mathbf{k}$, then
using the orthonormality property $\int d\Omega\ Y_{l^{\prime}m^{\prime}%
}^{\ast}(\mathbf{\hat{k}})Y_{lm}(\mathbf{\hat{k}})=\delta_{ll^{\prime}}%
\delta_{mm^{\prime}}$ of spherical harmonics we recover the expressions
(\ref{iso}) and (\ref{Cl2}) for the isotropic covariance matrix $\left\langle
a_{l^{\prime}m^{\prime}}^{\ast}a_{lm}\right\rangle $ and for the angular power
spectrum $C_{l}$ in terms of the isotropic primordial spectrum $\mathcal{P}%
_{\mathcal{R}}(k)$.

If instead we drop the assumption that $\left\langle \left\vert \mathcal{R}%
_{\mathbf{k}}\right\vert ^{2}\right\rangle $ is independent of the direction
of $\mathbf{k}$, we obtain an anisotropic primordial power spectrum%
\begin{equation}
\mathcal{P}_{\mathcal{R}}(k,\mathbf{\hat{k}})=(4\pi k^{3}/V)\left\langle
\left\vert \mathcal{R}_{\mathbf{k}}\right\vert ^{2}\right\rangle
\end{equation}
which depends on the (unit vector) direction $\mathbf{\hat{k}}$ in
$\mathbf{k}$-space as well as on the magnitude $k\equiv\left\vert
\mathbf{k}\right\vert $. This possibility has been discussed by several
authors \cite{AP06,ACW07,PK07,MEC11,MS15}, who have considered power spectra
of the form%
\begin{equation}
\mathcal{P}_{\mathcal{R}}(k,\mathbf{\hat{k}})=\mathcal{P}_{\mathcal{R}%
}(k)\left(  1+\sum_{LM}g_{LM}(k)Y_{LM}(\mathbf{\hat{k}})\right)  \ ,
\label{anisoPS}%
\end{equation}
where $\mathcal{P}_{\mathcal{R}}(k)$ is a standard isotropic spectrum and the
functions $g_{LM}(k)$ quantify scale-dependent anisotropies.

In terms of $\mathcal{P}_{\mathcal{R}}(k,\mathbf{\hat{k}})$ we have%
\begin{equation}
\left\langle a_{l^{\prime}m^{\prime}}^{\ast}a_{lm}\right\rangle =\frac{1}%
{2\pi^{2}}(-i)^{l^{\prime}}i^{l}\int_{0}^{\infty}\frac{dk}{k}\ \mathcal{T}%
(k,l^{\prime})\mathcal{T}(k,l)\int d\Omega\ \mathcal{P}_{\mathcal{R}%
}(k,\mathbf{\hat{k}})Y_{l^{\prime}m^{\prime}}^{\ast}(\mathbf{\hat{k}}%
)Y_{lm}(\mathbf{\hat{k}})\ . \label{covmat1'}%
\end{equation}
The consequences of an anisotropic spectrum will be explored below.

\subsection{Apparent anomalies in CMB data}

We now briefly consider the various anomalies that appear to exist in the CMB.

Data from the Planck satellite show evidence for an angular power deficit at
large scales. The 2013 release reported a deficit of 5--10\% in the region
$l\lesssim40$, with a statistical significance in the range 2.5--3$\sigma$
(depending on the estimator used) \cite{PlanckXV-2013}.\footnote{There were
already suggestions that data from the \textit{WMAP} satellite contained
anomalously low power at small $l$, but such claims were controversial
\cite{B11}.} The more recent 2015 release finds a similar deficit with a
slightly reduced significance \cite{Planck15-XI-PowerSpec}. A related anomaly
has been observed in the (temperature) two-point angular correlation function,
which at large scales is measured to be smaller than expected with a
statistical significance exceeding 3$\sigma$ \cite{CHSS13}. The reported
deficit could be a mere statistical fluctuation, or it might indicate a
genuine anomaly in the primordial power spectrum -- possibly from new physics.
In any case, it is important to develop theoretical models that predict a
low-$l$ deficit in order to better assess the nature and significance of this finding.

There are also significant hints of statistical anisotropy in various forms
and on different scales. For an extensive analysis and survey, see ref.
\cite{Planck15-XVI-IsoStats}. We briefly summarise the key points.

Signals of statistical anisotropy have been detected in the form of
hemispherical asymmetry, point-parity asymmetry (obtained by dividing the
temperature anisotropy into spherical harmonics with even and odd $l$-modes),
mirror-parity asymmetry (obtained from properties of the temperature
anisotropy under reflection with respect to a plane), and an improbable
alignment between the quadrupole ($l=2$) and octopole ($l=3$) modes. This last
feature was confirmed in the 2013 Planck data release \cite{PlanckXXIII-2013}.

Of particular interest to us are forms of statistical anisotropy that manifest
as correlations between $l$ and $l\pm1$ multipoles.

There is evidence for a dipolar modulation of the temperature anisotropy, of
the form \cite{GHHC05}%
\begin{equation}
\Delta T(\mathbf{\hat{n}})=(1+A\mathbf{\hat{p}}\cdot\mathbf{\hat{n}})\Delta
T_{\mathrm{iso}}(\mathbf{\hat{n}})\ ,\label{dipmod}%
\end{equation}
where $\Delta T_{\mathrm{iso}}$ is the temperature anisotropy of a
statistically isotropic sky, $\mathbf{\hat{p}}$ is a fixed unit vector and the
coefficient $A$ is measured to be $0.07\pm0.02$
\cite{PlanckXXIII-2013,Planck15-XVI-IsoStats}. However, significant power in
the dipole modulation seems to be limited to $l\simeq2-64$; for $l\gtrsim100$
there is no significant detection
\cite{PlanckXXIII-2013,Planck15-XVI-IsoStats}. The dipolar effects must
therefore depend on $l$; the simple temperature modulation (\ref{dipmod}) is
inadequate. One might also consider a possible quadrupolar modulation of the
primordial power spectrum (terms with $L=2$ in (\ref{anisoPS})), which would
induce correlations between $l$ and $l\pm2$ CMB multipoles. However, the
Planck team finds no evidence for such effects \cite{Planck15-XVI-IsoStats}.

There is intriguing evidence for statistical anisotropy even at small scales
\cite{PlanckXXIII-2013,Planck15-XVI-IsoStats}. The nature of the asymmetry is
not known, but it appears to be directional in some form. For the 2013 results
the relevant multipole range was $l=2-600$ \cite{PlanckXXIII-2013} (confirming
earlier findings in the WMAP data \cite{HBGEL09}). For the 2015 results, the
preferred dipolar directions associated with specific multipole ranges appear
to be correlated between large and small scales, where the correlations with
lower multipoles persist up to $l=1500$ \cite{Planck15-XVI-IsoStats}. Such
correlations are not expected for a statistically isotropic sky.

As emphasised by the Planck team, since many of the anomalies had already been
observed in the WMAP data it seems untenable that they could be caused by
systematic errors in these two independent experiments.

In this paper we focus on the statistical anisotropy as quantified by
$l-(l\pm1)$ multipole correlations.

\section{Unified mechanism for a power deficit with statistical anisotropy}

In this section we begin by summarising recent work that has yielded a
`quantum relaxation spectrum' given by an inverse-tangent power deficit. We
then discuss how this work can be generalised to allow for statistical
anisotropy -- in the Bunch-Davies vacuum, and in the primordial power
spectrum. Finally, we derive an expression for the anisotropic nonequilibrium
deficit in the limit of weak anisotropy. In this limit it is possible to
derive an explicit result that is fully determined (up to arbitrary constants)
by the inverse-tangent power deficit.

\subsection{Quantum relaxation spectrum}

In de Broglie-Bohm pilot-wave theory \cite{deB28,BV09,B52a,B52b,Holl93}, a
system with wave function $\psi(q,t)$ has a definite configuration $q(t)$
whose velocity $\dot{q}\equiv dq/dt$ is at all times determined by the general
formula%
\begin{equation}
\frac{dq}{dt}=\frac{j}{|\psi|^{2}}\ , \label{deB}%
\end{equation}
where $j=j\left[  \psi\right]  =j(q,t)$ is the Schr\"{o}dinger current
\cite{SV08}. The `pilot wave' $\psi(q,t)$ obeys the usual Schr\"{o}dinger
equation $i\partial\psi/\partial t=\hat{H}\psi$ (with $\hbar=1$) and is
defined in configuration space; it guides the motion of an individual system
and has no a priori connection with probability. Unlike in standard quantum
theory, an ensemble of systems with the same $\psi$ can have an arbitrary
distribution $\rho(q,t)$ of configurations. By construction, $\rho(q,t)$ obeys
the continuity equation%
\begin{equation}
\frac{\partial\rho}{\partial t}+\partial_{q}\cdot\left(  \rho\dot{q}\right)
=0\ . \label{cont}%
\end{equation}
The Schr\"{o}dinger equation for $\psi$ implies that $\left\vert
\psi\right\vert ^{2}$ also obeys (\ref{cont}). An initial distribution
$\rho(q,t_{i})=\left\vert \psi(q,t_{i})\right\vert ^{2}$ therefore evolves
into a final distribution $\rho(q,t)=\left\vert \psi(q,t)\right\vert ^{2}$.
This is the state of quantum equilibrium, for which we obtain the Born rule
and the usual predictions of quantum theory \cite{B52a,B52b}. For a
nonequilibrium ensemble ($\rho(q,t)\neq\left\vert \psi(q,t)\right\vert ^{2}$),
the statistical predictions generally disagree with those of quantum theory
\cite{AV91a,AV91b,AV92,AV07,PV06}.

In pilot-wave theory, quantum equilibrium may be understood to arise
dynamically. The process of relaxation or equilibration may be quantified by
an $H$-function $H=\int dq\ \rho\ln(\rho/\left\vert \psi\right\vert ^{2})$,
which obeys a coarse-graining $H$-theorem (where the minimum $H=0$ corresponds
to equilibrium $\rho=\left\vert \psi\right\vert ^{2}$) \cite{AV91a,AV92,AV01}.
If $\psi$ is a superposition of energy eigenstates, numerical simulations show
a rapid relaxation $\rho\rightarrow\left\vert \psi\right\vert ^{2}$ (on a
coarse-grained level) \cite{AV92,AV01,VW05,EC06,TRV12,SC12,ACV14}, with an
approximately exponential decay of the coarse-grained $H$-function
\cite{VW05,TRV12,ACV14}. It has been suggested that such relaxation took place
-- at least to a good approximation -- in the early universe
\cite{AV91a,AV91b,AV92,AV96,AV01,AV07,AV08,AV10}.

For the purposes of this paper, it suffices to consider the pilot-wave
dynamics of a free, minimally-coupled, real massless scalar field $\phi$ on an
expanding background with line element $d\tau^{2}=dt^{2}-a^{2}d\mathbf{x}^{2}%
$. Here $a=a(t)$ is the scale factor (with $c=1$). We have a classical
Lagrangian density $\mathcal{L}=\frac{1}{2}\sqrt{-g}g^{\mu\nu}\partial_{\mu
}\phi\partial_{\nu}\phi$, where $g_{\mu\nu}$ is the background metric. It is
convenient to work with Fourier components $\phi_{\mathbf{k}}=\frac{\sqrt{V}%
}{(2\pi)^{3/2}}\left(  q_{\mathbf{k}1}+iq_{\mathbf{k}2}\right)  $, where $V$
is a normalisation volume and the $q_{\mathbf{k}r}$'s ($r=1,2$) are real. The
field Hamiltonian is then a sum $H=\sum_{\mathbf{k}r}H_{\mathbf{k}r}$, where%
\begin{equation}
H_{\mathbf{k}r}=\frac{1}{2a^{3}}\pi_{\mathbf{k}r}^{2}+\frac{1}{2}%
ak^{2}q_{\mathbf{k}r}^{2}\ .
\end{equation}
Upon quantisation we obtain the Schr\"{o}dinger equation%
\begin{equation}
i\frac{\partial\Psi}{\partial t}=\sum_{\mathbf{k}r}\left(  -\frac{1}{2a^{3}%
}\frac{\partial^{2}}{\partial q_{\mathbf{k}r}^{2}}+\frac{1}{2}ak^{2}%
q_{\mathbf{k}r}^{2}\right)  \Psi
\end{equation}
for the wave functional $\Psi=\Psi\lbrack q_{\mathbf{k}r},t]$. We may then
identify the de Broglie velocities%
\begin{equation}
\frac{dq_{\mathbf{k}r}}{dt}=\frac{1}{a^{3}}\operatorname{Im}\frac{1}{\Psi
}\frac{\partial\Psi}{\partial q_{\mathbf{k}r}} \label{deB2}%
\end{equation}
for the configuration degrees of freedom $q_{\mathbf{k}r}$
\cite{AV07,AV08,AV10}. (This construction assumes a preferred foliation of
spacetime with time function $t$ and may be readily generalised to any
globally-hyperbolic spacetime \cite{AV04b,AV08a,AVbook}.)

We may consider the independent dynamics of an unentangled mode $\mathbf{k}$
with wave function $\psi_{\mathbf{k}}(q_{\mathbf{k}1},q_{\mathbf{k}2},t)$.
Dropping the index $\mathbf{k}$, we then have the Schr\"{o}dinger equation%
\begin{equation}
i\frac{\partial\psi}{\partial t}=\sum_{r=1,\ 2}\left(  -\frac{1}{2a^{3}%
}\partial_{r}^{2}+\frac{1}{2}ak^{2}q_{r}^{2}\right)  \psi
\end{equation}
for $\psi=\psi(q_{1},q_{2},t)$ and the de Broglie velocities%
\begin{equation}
\dot{q}_{r}=\frac{1}{a^{3}}\operatorname{Im}\frac{\partial_{r}\psi}{\psi}
\label{deB3}%
\end{equation}
for the configuration $(q_{1},q_{2})$ (with $\partial_{r}\equiv\partial
/\partial q_{r}$). The marginal distribution $\rho=\rho(q_{1},q_{2},t)$ obeys
the continuity equation%
\begin{equation}
\frac{\partial\rho}{\partial t}+\sum_{r=1,\ 2}\partial_{r}\left(  \rho\frac
{1}{a^{3}}\operatorname{Im}\frac{\partial_{r}\psi}{\psi}\right)  =0\ .
\end{equation}

Mathematically, the system is equivalent to a two-dimensional oscillator with
time-dependent mass $m=a^{3}$ and time-dependent angular frequency
$\omega=k/a$ \cite{AV07,AV08}. This is in turn equivalent to a standard
oscillator -- with constant mass and constant angular frequency -- but with
real time $t$ replaced by a `retarded time' $t_{\mathrm{ret}}(t,k)$
\cite{CV13}. Cosmological relaxation for a single field mode may then be
studied in terms of a standard oscillator.

Such studies have been carried out extensively for a radiation-dominated
expansion ($a\propto t^{1/2}$) \cite{CV13,CV15}. At short (sub-Hubble)
wavelengths, $t_{\mathrm{ret}}(t,k)\rightarrow t$ and we recover the usual
time evolution on Minkowski spacetime -- with rapid relaxation for a
superposition of excited states. At long (super-Hubble) wavelengths,
$t_{\mathrm{ret}}(t,k)<<t$ and relaxation is retarded (an effect which may
also be understood in terms of an upper bound on the mean displacement of the
trajectories \cite{AV08,AVbook}). Thus, if there was a radiation-dominated
pre-inflationary era, then at the beginning of inflation we may expect to find
relic nonequilibrium at long wavelengths \cite{AV10,CV13,CV15}.

No further relaxation takes place during inflation itself, because the de
Broglie-Bohm trajectories of the inflaton field are too simple -- assuming
that the Bunch-Davies vacuum is a good approximation \cite{AV07,AV10}. The
Bunch-Davies wave functional is a product $\Psi_{0}[q_{\mathbf{k}r}%
,t]=\prod\limits_{\mathbf{k}r}\psi_{\mathbf{k}r}(q_{\mathbf{k}r},t)$ of
contracting Gaussian packets and the de Broglie equation (\ref{deB2}) for the
trajectories $q_{\mathbf{k}r}(t)$ has a simple exact solution. It is found
that the time evolution of an arbitrary nonequilibrium distribution
$\rho_{\mathbf{k}r}(q_{\mathbf{k}r},t)$ shows the same overall contraction as
the time evolution of the equilibrium distribution, so that the ratio%
\begin{equation}
\xi(k)\equiv\frac{\left\langle |\phi_{\mathbf{k}}|^{2}\right\rangle
}{\left\langle |\phi_{\mathbf{k}}|^{2}\right\rangle _{\mathrm{QT}}}
\label{ksi}%
\end{equation}
of the nonequilibrium mean-square $\left\langle |\phi_{\mathbf{k}}%
|^{2}\right\rangle $ to the quantum-theoretical mean-square $\left\langle
|\phi_{\mathbf{k}}|^{2}\right\rangle _{\mathrm{QT}}$ is preserved in time.

Thus, in the approximation of a Bunch-Davies vacuum, if relic nonequilibrium
($\xi\neq1$) exists at the beginning of inflation it will be preserved during
the inflationary era and be transferred to larger lengthscales as physical
wavelengths $\lambda_{\mathrm{phys}}=a\lambda=a(2\pi/k)$ expand. The spectrum
of primordial curvature perturbations is then expected to take the form
\cite{AV07,AV10}%
\begin{equation}
\mathcal{P}_{\mathcal{R}}(k)=\mathcal{P}_{\mathcal{R}}^{\mathrm{QT}}%
(k)\xi(k)\ , \label{xi2}%
\end{equation}
where $\mathcal{P}_{\mathcal{R}}^{\mathrm{QT}}(k)$ is the quantum-theoretical
(equilibrium) spectrum. The function $\xi(k)$ quantifies the degree of
nonequilibrium as a function of $k$.

From measurements of the angular power spectrum $C_{l}$, we may set
observational bounds on $\xi(k)$ \cite{AV10}. Of greater interest is to make
predictions for $\xi(k)$ and to compare these with data. To this end, for a
single mode in a putative pre-inflationary era we considered initial wave
functions that are superpositions%
\begin{equation}
\psi(q_{1},q_{2},t_{i})=\frac{1}{\sqrt{M}}\sum_{n_{1}=0}^{\sqrt{M}-1}%
\sum_{n_{2}=0}^{\sqrt{M}-1}e^{i\theta_{n_{1}n_{2}}}\Phi_{n_{1}}(q_{1}%
)\Phi_{n_{2}}(q_{2}) \label{psi_i}%
\end{equation}
of $M$ energy eigenstates $\Phi_{n_{1}}\Phi_{n_{2}}$ of the initial
Hamiltonian, with coefficients $c_{n_{1}n_{2}}(t_{i})=(1/\sqrt{M}%
)e^{i\theta_{n_{1}n_{2}}}$ of equal amplitude and with randomly-chosen phases
$\theta_{n_{1}n_{2}}$. The wave function at time $t$ is then%
\begin{equation}
\psi(q_{1},q_{2},t)=\frac{1}{\sqrt{M}}\sum_{n_{1}=0}^{\sqrt{M}-1}\sum
_{n_{2}=0}^{\sqrt{M}-1}e^{i\theta_{n_{1}n_{2}}}\psi_{n_{1}}(q_{1}%
,t)\psi_{n_{2}}(q_{2},t)\ , \label{psi_t}%
\end{equation}
where $\psi_{n_{r}}(q_{r},t)$ is the time evolution of the initial wave
function $\psi_{n_{r}}(q_{r},t_{i})=\Phi_{n_{r}}(q_{r})$ under a formal
one-dimensional Schr\"{o}dinger equation with Hamiltonian $\hat{H}_{r}(t)$.
(The exact solution for $\psi_{n}(q,t)$ with $a\propto t^{1/2}$ was found in
ref. \cite{CV13}.) At time $t$ the equilibrium distribution is $\rho
_{\mathrm{QT}}(q_{1},q_{2},t)=|\psi(q_{1},q_{2},t)|^{2}$. In the simulations
reported in refs. \cite{CV13,CV15}, we assumed for simplicity an initial
nonequilibrium distribution%
\begin{equation}
\rho(q_{1},q_{2},t_{i})=|\Phi_{0}(q_{1})\Phi_{0}(q_{2})|^{2}=\frac{\omega
_{i}m_{i}}{\pi}e^{-m_{i}\omega_{i}q_{1}^{2}}e^{-m_{i}\omega_{i}q_{2}^{2}}\ .
\label{rho_i}%
\end{equation}
(This is just the equilibrium distribution for the ground state $\Phi
_{0}(q_{1})\Phi_{0}(q_{2})$.) The initial nonequilibrium width\ is smaller
than the initial equilibrium width. We calculated numerically the time
evolution $\rho(q_{1},q_{2},t)$ of the ensemble distribution up to a final
time $t_{f}$. This was done for varying values of $k$, as well as for varying
values of $M$ and $t_{f}$. In each case we compared the final nonequilibrium
width (that is, the width of $\rho(q_{1},q_{2},t_{f})$) with the final
equilibrium width (that is, the width of $|\psi(q_{1},q_{2},t_{f})|^{2}$) --
thereby yielding values of the ratio $\xi$ for varying values of $k$, $M$ and
$t_{f}$. (Here $\xi$ is defined as the ratio of the nonequilibrium and
equilibrium variances and not of the mean-squares as in (\ref{ksi}) which
holds only when the means vanish. Furthermore, before taking their ratio, the
respective variances were averaged over mixtures of initial wave functions
(\ref{psi_i}) with randomly-chosen phases. See ref. \cite{CV15} for details.)

For given $M$ and $t_{f}$, extensive numerical simulations for varying $k$
have yielded functions $\xi(k)$ that take the form of an inverse-tangent
\cite{CV15}. Specifically, in terms of the convenient variable%
\begin{equation}
x=2k/H_{0}\ , \label{x}%
\end{equation}
the numerical results show good fits to the curve%
\begin{equation}
\xi(k)=\tan^{-1}(\tilde{c}_{1}x+c_{2})-\frac{\pi}{2}+c_{3}\ , \label{numksi}%
\end{equation}
where $\tilde{c}_{1}$, $c_{2}$, $c_{3}$ are constant parameters that depend on
$M$ and $t_{f}$. Here $\tilde{c}_{1}=c_{1}RH_{0}/20\pi$, where the constant
$c_{1}$ is obtained from our simulations of a pre-inflationary era and the
factor $R\equiv a(t_{\mathrm{today}})/a(t_{f})$ is required to rescale our
wave numbers taking into account the spatial expansion from the end of
pre-inflation at time $t_{f}$ until today at time $t_{\mathrm{today}}$
\cite{CV15}. (Our numerical results for $c_{1}$, $c_{2}$, $c_{3}$ are listed
in tables I and II of ref. \cite{CV15}.) Note that for large $x$ we have
$\xi\rightarrow c_{3}$.

In ref. \cite{CV15} we considered $(\tilde{c}_{1},c_{2})\simeq(0.5,2.85)$ as
an illustrative example of values that are consistent with both our numerical
simulations and the approximate magnitude of the observed angular power
deficit, as well as with basic cosmological constraints on $R$. For
definiteness, as `fiducial' or illustrative values we may simply take%
\begin{equation}
(\tilde{c}_{1},c_{2})=(0.5,3) \label{fidc's}%
\end{equation}
(as will be used below). For $c_{3}$, in one case with $M=12$ we found
$c_{3}=0.95$. Taking this value for illustration, together with (\ref{fidc's}%
), in Figure 1 we plot the resulting $\xi$ as a function of $x$.%

\begin{figure}
[ptb]
\begin{center}
\includegraphics[width=0.7\textwidth]{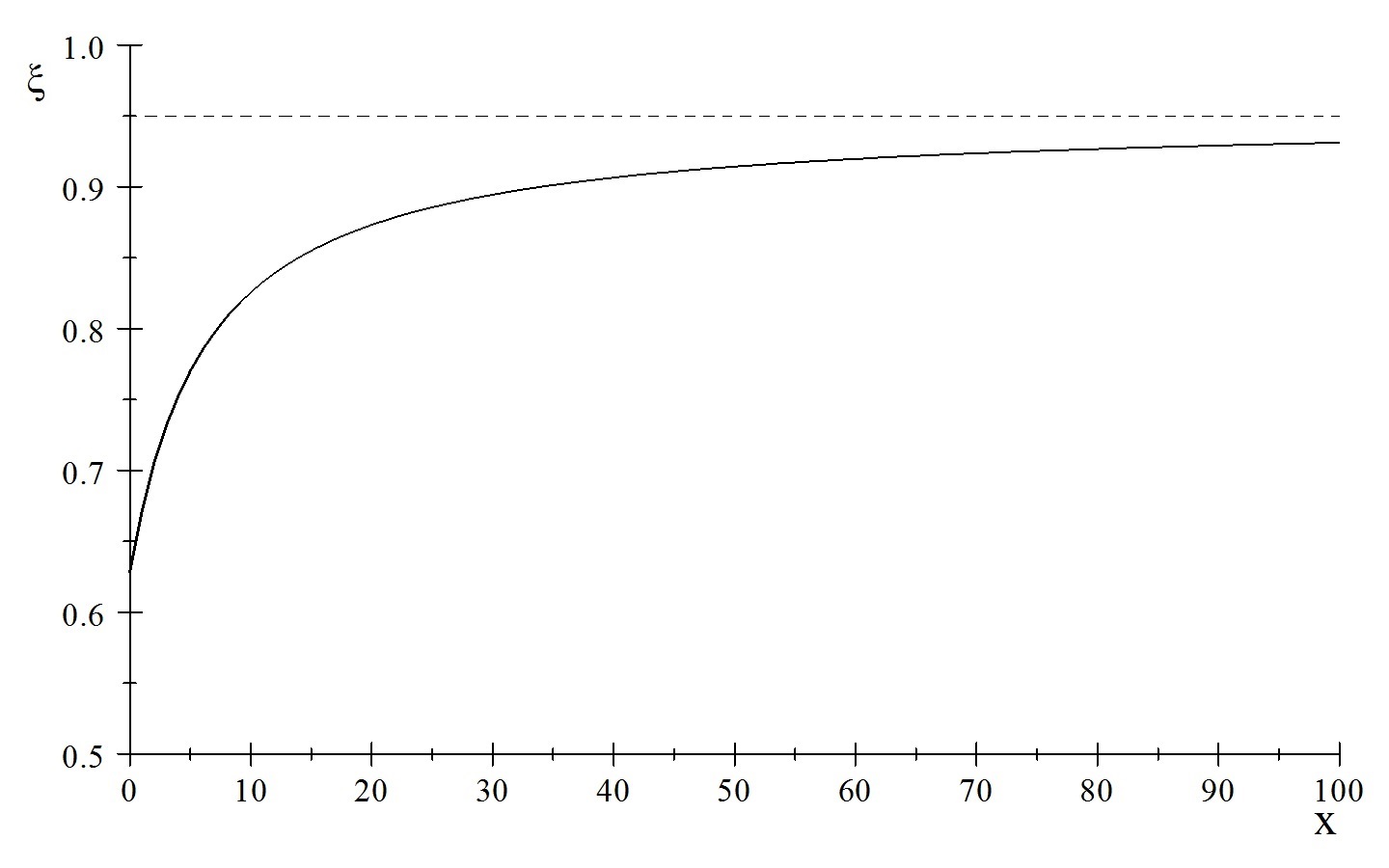}
\caption{Plot of the illustrative power deficit function $\xi=\tan
^{-1}(0.5x+3)-\frac{\pi}{2}+0.95$ (with $x=2k/H_{0}$). The dashed line shows
the asymptotic value $c_{3}=0.95$.}%
\end{center}
\end{figure}

The numerical results for $\xi(k)$ actually show oscillations around the curve
(\ref{numksi}) of magnitude $\lesssim10\%$ (see figure 3 of ref. \cite{CV15}).
To a first approximation we ignore the oscillations, whose contribution we
hope to consider in future work.

The overall shape of the curve (\ref{numksi}) shows the retardation or
suppression of relaxation at long wavelengths, with $\xi(k)$ decreasing for
smaller $k$. At short wavelengths, or large $k$, we might expect complete
relaxation -- in which case we should have $\xi\rightarrow1$. Instead, we have
$\xi\rightarrow c_{3}$ and the numerical results show that $c_{3}$ can differ
from $1$. Such a nonequilibrium `residue' at short wavelengths is equivalent
to an overall renormalisation of the power spectrum and is therefore by itself
observationally indistinguishable from a compensating shift in other
cosmological parameters \cite{CV15,PVV15}.

The numerical results of ref. \cite{CV15} indicate that the large-$k$ residue
$c_{3}$ approaches $1$ for large $M$. As expected, we find equilibrium to good
accuracy ($\xi\simeq1$) in a limiting regime with both short wavelengths and
large numbers of modes. However, for small $M$ the residue $c_{3}$ is not
quite equal to $1$. Even at very short wavelengths, for small $M$ relaxation
is unlikely to occur completely because the trajectories are unlikely to fully
explore the configuration space (depending on the initial phases in the
wavefunction) \cite{ACV14}. In our scenario there is a preference for small
$M$ during pre-inflation, because after the transition to inflation a
significant number of excitations above the vacuum are likely to cause a
back-reaction problem \cite{BM13}. Thus, in a realistic working scenario
$c_{3}$ is likely to differ slightly from $1$.

The aim of ref. \cite{CV15} was to obtain a general signature of cosmological
quantum relaxation that is as far as possible independent of details of the
pre-inflationary era. In more recent work \cite{CV15b}, the generality of the
inverse-tangent result (\ref{numksi}) has been extended in two ways. First, we
have studied initial nonequilibrium distributions that are more complicated
than the initial `vacuum' Gaussian (\ref{rho_i}). Second, we have included a
period of exponential expansion after the radiation-dominated phase. In both
cases, we still obtain a function $\xi(k)$ of the form (\ref{numksi}). We may
therefore reasonably expect a deficit of the form (\ref{numksi}) in any
scenario involving cosmological quantum relaxation. Note that while we usually
find $c_{3}<1$, sometimes we find $c_{3}>1$ \cite{CV15b}.

In all of our numerical simulations of quantum relaxation, we assume initial
distributions $\rho(q_{1},q_{2},t_{i})$ whose width is smaller than the
initial equilibrium width (that is, smaller than the width of $|\psi
(q_{1},q_{2},t_{i})|^{2}$). In principle, this assumption{} could be violated
by the actual initial conditions of our universe. However, we make this
assumption partly for simplicity and partly on heuristic grounds: if quantum
noise has a dynamical origin, it seems natural to assume initial conditions
with a subquantum statistical spread (so that the initial state contains less
noise than an ordinary quantum state). In any case, some assumptions need to
made about the initial conditions in order to make any predictions at all.
Another assumption we make is that $\rho(q_{1},q_{2},t_{i})$ does not possess
any fine-grained micro-structure; that is, we assume that $\rho(q_{1}%
,q_{2},t_{i})$ is approximately constant over the size of our coarse-graining
cells (used to calculate the decreasing coarse-grained $H$-function). Some
such assumption is always required in order to obtain relaxation in the
statistical mechanics of a time-reversal invariant theory.\footnote{For
discussions of this point in pilot-wave theory, see refs.
\cite{AV92,AV96,AV01,VW05}.} Our assumptions about the initial state are
guided by simplicity and heuristic arguments, but in the end their
justification rests on how well the resulting predictions compare with observation.

In the reported numerical studies, the scalar field $\phi$ is regarded as
representative of whatever generic fields may have been present in a
pre-inflationary era. The relation between our field $\phi$ and the later
inflaton field is not specified. This is a significant gap in our proposed
scenario, which remains to be filled. We simply assume that the simulated
deficit function $\xi(k)$ at the end of pre-inflation more or less survives
the transition to inflation in the sense that the same correction appears in
the inflationary spectrum \cite{AV10,CV13,CV15}. A more complete understanding
must await the development of a model of the transition from pre-inflation to
inflation -- which may for example involve symmetry breaking and associated
field redefinitions. It is conceivable that the transition will in fact
significantly change the functional form of the predicted deficit. Pending a
proper treatment, we assume that this is not the case. With this caveat, we
may say that the deficit function (\ref{numksi}) is a robust prediction of
cosmological quantum relaxation.

In ref. \cite{CV15} we also studied relaxation for the phases of our
primordial scalar field components $\phi_{\mathbf{k}}$, with a view to
understanding statistical anisotropy. We found that anomalous phases are
likely to exist at scales comparable to that of the power deficit. However,
the connection with the observed anisotropy of the CMB was not clearly
established as there is no simple relationship between the phases of
primordial perturbations $\mathcal{R}_{\mathbf{k}}\propto\phi_{\mathbf{k}}$
and the phases of CMB harmonic coefficients $a_{lm}$. Here we follow a simpler
and more direct route to statistical anisotropy by considering primordial
perturbations whose variance depends on the wave vector direction
$\mathbf{\hat{k}}$ (a suggestion that was in fact briefly made at the end of
ref. \cite{CV13}).

\subsection{Statistical anisotropy in the Bunch-Davies vacuum}

Let us consider how statistical anisotropy can exist in the Bunch-Davies
vacuum. In conventional quantum field theory this would be a contradiction in
terms, but it is perfectly possible in pilot-wave field theory. To show this,
we first recall some basic properties of the Bunch-Davies vacuum $\left\vert
0\right\rangle $ for a free scalar field $\phi$ on de Sitter spacetime
($a\propto e^{Ht}$) \cite{AV07,AV10}.

The Bunch-Davies wave functional $\Psi_{0}[q_{\mathbf{k}r},t]=\langle
q_{\mathbf{k}r}\left\vert 0\right\rangle $ may be calculated from the defining
equation $\hat{a}_{\mathbf{k}}\left\vert 0\right\rangle =0$, where $\hat
{a}_{\mathbf{k}}$ is an appropriate annihilation operator \cite{AV10}. The
result is a product $\Psi_{0}[q_{\mathbf{k}r},t]=\prod\limits_{\mathbf{k}%
r}\psi_{\mathbf{k}r}(q_{\mathbf{k}r},t)$ of wave functions $\psi_{\mathbf{k}%
r}=\left\vert \psi_{\mathbf{k}r}\right\vert e^{is_{\mathbf{k}r}}$, where%
\begin{equation}
\left\vert \psi_{\mathbf{k}r}\right\vert ^{2}=\frac{1}{\sqrt{2\pi\Delta
_{k}^{2}}}e^{-q_{\mathbf{k}r}^{2}/2\Delta_{k}^{2}} \label{psi2}%
\end{equation}
is a Gaussian of contracting squared width%
\begin{equation}
\Delta_{k}^{2}=\frac{H^{2}}{2k^{3}}\left(  1+\frac{k^{2}}{H^{2}a^{2}}\right)
\label{D2}%
\end{equation}
and the phase $s_{\mathbf{k}r}$ is given by%
\begin{equation}
s_{\mathbf{k}r}=-\frac{ak^{2}q_{\mathbf{k}r}^{2}}{2H(1+k^{2}/H^{2}a^{2}%
)}+h(t)\ , \label{s}%
\end{equation}
where%
\begin{equation}
h(t)=\frac{1}{2}\left(  \frac{k}{Ha}-\tan^{-1}\left(  \frac{k}{Ha}\right)
\right)  \ .
\end{equation}
The wave function $\psi_{\mathbf{k}r}=\left\vert \psi_{\mathbf{k}r}\right\vert
e^{is_{\mathbf{k}r}}$ satisfies the Schr\"{o}dinger equation for a single mode
$\mathbf{k}r$, and in the limit $H\longrightarrow0$, $a\rightarrow1$ reduces
to the wave function of the Minkowski vacuum $\psi_{\mathbf{k}r}\propto
e^{-\frac{1}{2}kq_{\mathbf{k}r}^{2}}e^{-i\frac{1}{2}kt}$. Note that the width
$\Delta_{\mathbf{k}r}=\Delta_{k}$ depends on the magnitude $k=\left\vert
\mathbf{k}\right\vert $ of the wave vector but not on its direction
$\mathbf{\hat{k}}$.

In terms of conformal time $\eta=-1/Ha$ (ranging from $-\infty$ to $0$), the
de Broglie equation of motion for $q_{\mathbf{k}r}$ reads%
\begin{equation}
\frac{dq_{\mathbf{k}r}}{d\eta}=\frac{k^{2}\eta q_{\mathbf{k}r}}{1+k^{2}%
\eta^{2}}\ , \label{dqdeta}%
\end{equation}
with the solution \cite{AV07,AV10}%
\begin{equation}
q_{\mathbf{k}r}(\eta)=q_{\mathbf{k}r}(0)\sqrt{1+k^{2}\eta^{2}} \label{traj}%
\end{equation}
(in terms of the asymptotic values at $\eta=0$). It is also easy to obtain the
time evolution of an arbitrary initial probability distribution. For
simplicity we may assume that the initial distribution $P[q_{\mathbf{k}r}%
,\eta_{i}]$ factorises across modes. For a single mode, the marginal
distribution $\rho_{\mathbf{k}r}(q_{\mathbf{k}r},\eta)$ (generally
$\neq\left\vert \psi_{\mathbf{k}r}(q_{\mathbf{k}r},\eta)\right\vert ^{2}$)
satisfies the continuity equation%
\begin{equation}
\frac{\partial\rho_{\mathbf{k}r}}{\partial\eta}+\frac{\partial}{\partial
q_{\mathbf{k}r}}\left(  \rho_{\mathbf{k}r}\frac{dq_{\mathbf{k}r}}{d\eta
}\right)  =0 \label{Contrhoeta}%
\end{equation}
(with the velocity field (\ref{dqdeta})). This has the general solution
\cite{AV10}%
\begin{equation}
\rho_{\mathbf{k}r}(q_{\mathbf{k}r},\eta)=\frac{1}{\sqrt{1+k^{2}\eta^{2}}}%
\rho_{\mathbf{k}r}(q_{\mathbf{k}r}/\sqrt{1+k^{2}\eta^{2}},0) \label{rhoeta}%
\end{equation}
for any given $\rho_{\mathbf{k}r}(q_{\mathbf{k}r},0)$ (again writing formally
in terms of asymptotic values at $\eta=0$).

At times $\eta<0$, $\left\vert \psi_{\mathbf{k}r}\right\vert ^{2}$ is a
contracting Gaussian of width%
\begin{equation}
\Delta_{k}(\eta)=\Delta_{k}(0)\sqrt{1+k^{2}\eta^{2}}%
\end{equation}
(with $\Delta_{k}(0)=\sqrt{H^{2}/2k^{3}}$) while $\rho_{\mathbf{k}r}$ is a
contracting arbitrary distribution of width%
\begin{equation}
D_{\mathbf{k}r}(\eta)=D_{\mathbf{k}r}(0)\sqrt{1+k^{2}\eta^{2}}%
\end{equation}
(with arbitrary $D_{\mathbf{k}r}(0)$). Both distributions contract by the same
overall factor and so their widths remain in a constant ratio.

Now, in earlier studies \cite{AV07,AV10,CV13} it was assumed for simplicity
that $D_{\mathbf{k}r}(\eta)=D_{k}(\eta)$. The nonequilibrium width
$D_{\mathbf{k}r}$ was taken to be independent of the direction of $\mathbf{k}$
(as well as independent of $r$), as is always the case for the equilibrium
width $\Delta_{\mathbf{k}r}=\Delta_{k}$. We then defined the deficit function
as the ratio $\xi(k)=D_{k}^{2}/\Delta_{k}^{2}$, which reduces to (\ref{ksi})
when the distributions have zero mean. (The mean is necessarily zero for the
equilibrium vacuum, and for simplicity we assume it to be zero for the
nonequilibrium vacuum as well.) However, for a general nonequilibrium state
there is no physical reason -- other than simplicity -- for why the width
$D_{\mathbf{k}r}$ should not also depend on the wave vector direction
$\mathbf{\hat{k}}$ (as well as on $r$).

We may then define a direction-dependent deficit function $\xi=\xi
(k,\mathbf{\hat{k}})$ by the (time-independent) ratio%
\begin{equation}
\xi(k,\mathbf{\hat{k}})\equiv\frac{D_{\mathbf{k}1}^{2}+D_{\mathbf{k}2}^{2}%
}{\Delta_{\mathbf{k}1}^{2}+\Delta_{\mathbf{k}2}^{2}}\ ,
\end{equation}
where $D_{\mathbf{k}r}^{2}=\left\langle q_{\mathbf{k}r}^{2}\right\rangle
-\left\langle q_{\mathbf{k}r}\right\rangle ^{2}$ and $\Delta_{\mathbf{k}r}%
^{2}=\left\langle q_{\mathbf{k}r}^{2}\right\rangle _{\mathrm{QT}}-\left\langle
q_{\mathbf{k}r}\right\rangle _{\mathrm{QT}}^{2}$. The equilibrium variance
$\Delta_{\mathbf{k}r}^{2}=\Delta_{k}^{2}$ is in fact independent of
$\mathbf{\hat{k}}$ and $r$. We also have $\left\langle q_{\mathbf{k}%
r}\right\rangle _{\mathrm{QT}}=0$. If we assume that the nonequilibrium mean
also vanishes, $\left\langle q_{\mathbf{k}r}\right\rangle =0$, we then have%
\begin{equation}
\xi(k,\mathbf{\hat{k}})=\frac{\left\langle q_{\mathbf{k}1}^{2}\right\rangle
+\left\langle q_{\mathbf{k}2}^{2}\right\rangle }{\left\langle q_{\mathbf{k}%
1}^{2}\right\rangle _{\mathrm{QT}}+\left\langle q_{\mathbf{k}2}^{2}%
\right\rangle _{\mathrm{QT}}}=\frac{\left\langle |\phi_{\mathbf{k}}%
|^{2}\right\rangle }{\left\langle |\phi_{\mathbf{k}}|^{2}\right\rangle
_{\mathrm{QT}}}\ .
\end{equation}
Thus our direction-dependent mean-square perturbation may be written as%
\begin{equation}
\left\langle |\phi_{\mathbf{k}}|^{2}\right\rangle =\left\langle |\phi
_{\mathbf{k}}|^{2}\right\rangle _{\mathrm{QT}}\xi(k,\mathbf{\hat{k}})\ .
\label{BDaniso}%
\end{equation}
Statistical isotropy is violated -- even though the wave functional itself is
still that of the Bunch-Davies vacuum.

\subsection{Derivation of the anisotropic deficit function}

If we take the field perturbations $\phi_{\mathbf{k}}$ to generate primordial
curvature perturbations $\mathcal{R}_{\mathbf{k}}\propto\phi_{\mathbf{k}}$,
then from (\ref{BDaniso}) we will have a direction-dependent mean-square%
\begin{equation}
\left\langle \left\vert \mathcal{R}_{\mathbf{k}}\right\vert ^{2}\right\rangle
=\left\langle \left\vert \mathcal{R}_{\mathbf{k}}\right\vert ^{2}\right\rangle
_{\mathrm{QT}}\xi(k,\mathbf{\hat{k}})
\end{equation}
and an anisotropic primordial power spectrum%
\begin{equation}
\mathcal{P}_{\mathcal{R}}(k,\mathbf{\hat{k}})=\mathcal{P}_{\mathcal{R}%
}^{\mathrm{QT}}(k)\xi(k,\mathbf{\hat{k}})\ . \label{PSaniso}%
\end{equation}
The isotropic equilibrium spectrum $\mathcal{P}_{\mathcal{R}}^{\mathrm{QT}%
}(k)$ is modulated by the anisotropic deficit function $\xi(k,\mathbf{\hat{k}%
})$.

What could be the physical origin of such an anisotropic deficit? The
equilibrium spectrum $\mathcal{P}_{\mathcal{R}}^{\mathrm{QT}}(k)$ is
necessarily isotropic because the quantum width $\Delta_{\mathbf{k}r}%
=\Delta_{k}$ of inflationary vacuum modes is independent of $\mathbf{\hat{k}}%
$. However, if quantum nonequilibrium existed in a pre-inflationary phase we
may expect it to be anisotropic since the nonequilibrium width $D_{\mathbf{k}%
r}$ of pre-inflationary field modes will generally depend on $\mathbf{\hat{k}%
}$. Furthermore, relaxation suppression at long wavelengths will preserve the
anisotropy (as well as any power deficit) for small $k$ while efficient
relaxation at short wavelengths will drive the field modes close to the
quantum isotropic spectrum (with the usual power) for large $k$. Thus the
dynamics of pilot-wave field theory on expanding space provides a natural
mechanism whereby both a power deficit and a statistical anisotropy can be
generated at large scales while approximately yielding the standard
predictions at small scales. In fact, as we shall see, at small scales we
expect to find an approximately scale-free spectrum but with residual
statistical anisotropies. It is this possibility -- of a unified explanation
for the large-scale deficit and the large-scale anisotropy, together with the
small-scale anisotropy -- that we wish to explore and develop in this paper.

A general anisotropic power spectrum $\mathcal{P}_{\mathcal{R}}(k,\mathbf{\hat
{k}})$ may be written in the form (\ref{anisoPS}) with%
\begin{equation}
\xi(k,\mathbf{\hat{k}})=1+\sum_{LM}g_{LM}(k)Y_{LM}(\mathbf{\hat{k}})\ ,
\label{ksiexpn}%
\end{equation}
where $g_{LM}(k)$ are scale-dependent harmonic coefficients in $\mathbf{k}%
$-space. In a general phenomenology, the $g_{LM}(k)$ would be arbitrary
functions of $k$. However, our model with cosmological quantum relaxation
implies a specific $k$-dependence (in the limit of weak anisotropy).

The isotropic nonequilibrium deficit function (\ref{numksi}) was obtained from
numerical simulations \cite{CV15}. These were carried out for single field
modes of variable wave number $k$ and fixed wave vector direction
$\mathbf{\hat{k}}$. For each $k$ an average was taken over an ensemble of wave
functions with randomly-chosen initial phases. While the dynamics depends on
$k$, it is blind to the direction $\mathbf{\hat{k}}$. We may then imagine
carrying out a generalised simulation for single field modes of variable $k$
and $\mathbf{\hat{k}}$, where the initial conditions -- in particular the
initial nonequilibrium distribution $\rho(q_{\mathbf{k}1},q_{\mathbf{k}%
2},t_{i})$ and its width -- are allowed to depend on $\mathbf{\hat{k}}$. For
each mode with given $k$ and $\mathbf{\hat{k}}$, we again take an average over
an ensemble of wave functions with randomly-chosen initial phases. Now
different initial nonequilibrium states will in general yield different
coefficients $\tilde{c}_{1}$, $c_{2}$, $c_{3}$ in the function $\xi(k)$
generated by the simulations. (Some examples are in fact found in ref.
\cite{CV15b}.) Thus, if field modes with different wave vector directions
$\mathbf{\hat{k}}$ have different initial nonequilibrium distributions and
widths, then in effect the generated coefficients $\tilde{c}_{1}$, $c_{2}$,
$c_{3}$ will depend on $\mathbf{\hat{k}}$ and we may consider functions
$\tilde{c}_{1}(\mathbf{\hat{k}})$, $c_{2}(\mathbf{\hat{k}})$, $c_{3}%
(\mathbf{\hat{k}})$. Let us write these as%
\[
\tilde{c}_{1}(\mathbf{\hat{k}})=\tilde{c}_{1}+\Delta_{1}(\mathbf{\hat{k}%
})\ ,\ \ \ c_{2}(\mathbf{\hat{k}})=c_{2}+\Delta_{2}(\mathbf{\hat{k}%
})\ ,\ \ \ c_{3}(\mathbf{\hat{k}})=c_{3}+\Delta_{3}(\mathbf{\hat{k}})\ ,
\]
with the $\Delta_{i}(\mathbf{\hat{k}})$ expanded as%
\begin{equation}
\Delta_{i}(\mathbf{\hat{k}})=\sum_{L\geq1,M}\Delta_{LM}^{i}Y_{LM}%
(\mathbf{\hat{k}}) \label{expns'}%
\end{equation}
(for $i=1,2,3$, where the coefficients $\Delta_{LM}^{i}$ are independent of
$k$). In this way the isotropic function (\ref{numksi}) is generalised to the
anisotropic function%
\[
\xi(k,\mathbf{\hat{k}})=\tan^{-1}\left[  \left(  \tilde{c}_{1}+\Delta
_{1}(\mathbf{\hat{k}})\right)  x+c_{2}+\Delta_{2}(\mathbf{\hat{k}})\right]
-\frac{\pi}{2}+c_{3}+\Delta_{3}(\mathbf{\hat{k}})\ .
\]

In the limit of weak anisotropy we may take the $\Delta_{i}(\mathbf{\hat{k}})$
to be small. Expanding to lowest order in $\Delta_{i}(\mathbf{\hat{k}})$ and
defining%
\begin{equation}
f_{1}(x)\equiv\frac{\partial\xi(k)}{\partial\tilde{c}_{1}}\ ,\ \ \ f_{2}%
(x)\equiv\frac{\partial\xi(k)}{\partial c_{2}}\ ,\ \ \ f_{3}(x)\equiv
\frac{\partial\xi(k)}{\partial c_{3}} \label{fns_defn}%
\end{equation}
(again with $x=2k/H_{0}$), we have%
\begin{equation}
\xi(k,\mathbf{\hat{k}})=\xi(k)+\sum_{i=1}^{3}f_{i}(x)\Delta_{i}(\mathbf{\hat
{k}})\ , \label{ksi_aniso_0}%
\end{equation}
where from (\ref{numksi}) we find the functions%
\begin{equation}
f_{1}(x)=\frac{x}{1+(\tilde{c}_{1}x+c_{2})^{2}}\ ,\ \ \ \ f_{2}(x)=\frac
{1}{1+(\tilde{c}_{1}x+c_{2})^{2}}\ ,\ \ \ \ f_{3}(x)=1\ . \label{fns}%
\end{equation}
The anisotropic part of $\xi(k,\mathbf{\hat{k}})$ depends on $\tilde{c}_{1}$
and $c_{2}$ but is independent of $c_{3}$. Below we shall consider the
functions $f_{i}(x)$ with the illustrative values (\ref{fidc's}) for
$\tilde{c}_{1}$, $c_{2}$.

Using the expansions (\ref{expns'}), we obtain an expression for the
anisotropic deficit function%
\begin{equation}
\xi(k,\mathbf{\hat{k}})=\xi(k)+\sum_{L\geq1,M}\left(  \sum_{i=1}^{3}%
f_{i}(x)\Delta_{LM}^{i}\right)  Y_{LM}(\mathbf{\hat{k}}) \label{ksi_aniso}%
\end{equation}
in terms of unknown coefficients $\Delta_{LM}^{i}$ but with known scaling
functions $f_{i}(x)$.

The result (\ref{ksi_aniso}) may be compared with the general expression
(\ref{ksiexpn}). We have $Y_{00}=1/\sqrt{4\pi}$ and so%
\begin{equation}
g_{00}(k)=-\sqrt{4\pi}\left(  1-\xi(k)\right)  \ . \label{g_00}%
\end{equation}
For $L\geq1$ we have%
\begin{equation}
g_{LM}(k)=\sum_{i=1}^{3}f_{i}(x)\Delta_{LM}^{i}\ . \label{g_LM}%
\end{equation}
The third term ($i=3$) is independent of $k$, whereas the first two terms
scale with $k$ and tend to zero for large $k$.

Note that for (\ref{ksi_aniso}) we have $\frac{1}{4\pi}\int d\Omega
\ \xi(k,\mathbf{\hat{k}})=\xi(k)$. The angular average of $\xi(k,\mathbf{\hat
{k}})$ is equal to the average isotropic result $\xi(k)$ obtained from the
numerical simulations.

We may now explore the implications of the anisotropic deficit function
(\ref{ksi_aniso_0}) or (\ref{ksi_aniso}).

In the expansions (\ref{expns'}) the reality of $\Delta_{i}(\mathbf{\hat{k}})$
implies that%
\begin{equation}
(\Delta_{LM}^{i})^{\ast}=(-1)^{M}\Delta_{L(-M)}^{i}\ ,
\end{equation}
where $Y_{LM}^{\ast}=(-1)^{M}Y_{L(-M)}$. Taking angular coordinates
$(\theta,\phi)$ in $\mathbf{k}$-space, with $Y_{10}=\sqrt{3/4\pi}\cos\theta$
and $Y_{11}=-\sqrt{3/8\pi}\sin\theta.e^{i\phi}$, the first few terms in the
expansions take the form%
\begin{equation}
\Delta_{i}(\mathbf{\hat{k}})=\Delta_{10}^{i}\sqrt{3/4\pi}\cos\theta
-\Delta_{11}^{i}\sqrt{3/8\pi}e^{i\phi}\sin\theta-(\Delta_{11}^{i})^{\ast}%
\sqrt{3/8\pi}e^{-i\phi}\sin\theta+...\ ,\label{delta_expn2}%
\end{equation}
where the $\Delta_{10}^{i}$ are real.

We might reasonably assume that the first few terms will be the most
important. For simplicity we shall consider only the lowest ($L=1$) terms. As
discussed further in Section 6, in our scenario the initial anisotropy is in
principle arbitrary and it is arguably natural to expect the lowest ($L=1$)
terms to dominate.

\section{Anisotropic corrections to the covariance matrix}

In Section 2.3 we expressed the covariance matrix $\left\langle a_{l^{\prime
}m^{\prime}}^{\ast}a_{lm}\right\rangle $ in terms of a general anisotropic
power spectrum $\mathcal{P}_{\mathcal{R}}(k,\mathbf{\hat{k}})$. The result
(\ref{covmat1'}) is valid as long as the primordial perturbations
$\mathcal{R}_{\mathbf{k}}$ are statistically homogeneous (as we assume
throughout). For $\mathcal{P}_{\mathcal{R}}(k,\mathbf{\hat{k}})=\mathcal{P}%
_{\mathcal{R}}^{\mathrm{QT}}(k)\xi(k,\mathbf{\hat{k}})$ we then have%
\begin{equation}
\left\langle a_{l^{\prime}m^{\prime}}^{\ast}a_{lm}\right\rangle =\frac
{(-i)^{l^{\prime}}i^{l}}{2\pi^{2}}\int_{0}^{\infty}\frac{dk}{k}\ \mathcal{T}%
(k,l^{\prime})\mathcal{T}(k,l)\mathcal{P}_{\mathcal{R}}^{\mathrm{QT}}%
(k)\chi_{l^{\prime}m^{\prime}lm}(k)\ , \label{covmat2}%
\end{equation}
where we have defined%
\begin{equation}
\chi_{l^{\prime}m^{\prime}lm}(k)\equiv\int d\Omega\ \xi(k,\mathbf{\hat{k}%
})Y_{l^{\prime}m^{\prime}}^{\ast}(\mathbf{\hat{k}})Y_{lm}(\mathbf{\hat{k}})\ .
\label{Chi}%
\end{equation}

In quantum equilibrium $\xi(k,\mathbf{\hat{k}})=1$ and the quantities
$\chi_{l^{\prime}m^{\prime}lm}(k)$ take the quantum-theoretical values%
\begin{equation}
\chi_{l^{\prime}m^{\prime}lm}^{\mathrm{QT}}(k)=\delta_{ll^{\prime}}%
\delta_{mm^{\prime}}\ .
\end{equation}
We then recover the equilibrium covariance matrix%
\begin{equation}
\left\langle a_{l^{\prime}m^{\prime}}^{\ast}a_{lm}\right\rangle _{\mathrm{QT}%
}=\delta_{ll^{\prime}}\delta_{mm^{\prime}}C_{l}^{\mathrm{QT}} \label{iso'}%
\end{equation}
and the equilibrium angular power spectrum%
\begin{equation}
C_{l}^{\mathrm{QT}}=\frac{1}{2\pi^{2}}\int_{0}^{\infty}\frac{dk}%
{k}\ \mathcal{T}^{2}(k,l)\mathcal{P}_{\mathcal{R}}^{\mathrm{QT}}(k)
\label{Cl2'}%
\end{equation}
in terms of the (isotropic)\ equilibrium primordial spectrum $\mathcal{P}%
_{\mathcal{R}}^{\mathrm{QT}}(k)$. The expressions (\ref{iso'}) and
(\ref{Cl2'}) are usually associated with fundamental statistical isotropy.
Here we regard them as peculiar to quantum equilibrium -- which happens to
yield statistical isotropy in the standard inflationary vacuum.

If instead $\xi(k,\mathbf{\hat{k}})$ depends on the direction vector
$\mathbf{\hat{k}}$, then isotropy is generally broken and the nonequilibrium
covariance matrix (\ref{covmat2}) takes a more general form.

\subsection{Covariance matrix with the anisotropic deficit function}

Let us consider our anisotropic deficit function in the general form
(\ref{ksi_aniso_0}), for which the quantities (\ref{Chi}) become%
\[
\chi_{l^{\prime}m^{\prime}lm}(k)\equiv\xi(k)\delta_{ll^{\prime}}%
\delta_{mm^{\prime}}+\sum_{i=1}^{3}f_{i}(x)I_{i}(l^{\prime},m^{\prime};l,m)
\]
where%
\begin{equation}
I_{i}(l^{\prime},m^{\prime};l,m)\equiv\int d\Omega\ \Delta_{i}(\mathbf{\hat
{k}})Y_{l^{\prime}m^{\prime}}^{\ast}(\mathbf{\hat{k}})Y_{lm}(\mathbf{\hat{k}%
})\ . \label{I_i}%
\end{equation}
Because $\Delta_{i}$ is real we must have%
\begin{equation}
I_{i}^{\ast}(l^{\prime},m^{\prime};l,m)=I_{i}(l,m;l^{\prime},m^{\prime})\ .
\end{equation}

Our nonequilibrium covariance matrix (\ref{covmat2}) then becomes%
\begin{equation}
\left\langle a_{l^{\prime}m^{\prime}}^{\ast}a_{lm}\right\rangle =\delta
_{ll^{\prime}}\delta_{mm^{\prime}}C_{l}+\frac{(-i)^{l^{\prime}}i^{l}}{2\pi
^{2}}\sum_{i=1}^{3}I_{i}(l^{\prime},m^{\prime};l,m)F_{i}(l^{\prime},l)
\label{covmat3}%
\end{equation}
where $C_{l}$ is now the corrected angular power spectrum%
\begin{equation}
C_{l}=\frac{1}{2\pi^{2}}\int_{0}^{\infty}\frac{dk}{k}\ \mathcal{T}%
^{2}(k,l)\mathcal{P}_{\mathcal{R}}^{\mathrm{QT}}(k)\xi(k) \label{C_l_corr}%
\end{equation}
(written in terms of an isotropic primordial spectrum $\mathcal{P}%
_{\mathcal{R}}(k)=\mathcal{P}_{\mathcal{R}}^{\mathrm{QT}}(k)\xi(k)$ with an
isotropic deficit function $\xi(k)$), and where for $i=1,2,3$ we have defined
the integrals%
\begin{equation}
F_{i}(l^{\prime},l)\equiv\int_{0}^{\infty}\frac{dk}{k}\ \mathcal{T}%
(k,l^{\prime})\mathcal{T}(k,l)\mathcal{P}_{\mathcal{R}}^{\mathrm{QT}}%
(k)f_{i}(x)\ . \label{F_i}%
\end{equation}
Note that the anisotropic part of $\left\langle a_{l^{\prime}m^{\prime}}%
^{\ast}a_{lm}\right\rangle $ depends on $\tilde{c}_{1}$ and $c_{2}$ but is
independent of $c_{3}$.

We are particularly interested in the off-diagonal matrix elements. For
$l^{\prime}m^{\prime}\neq lm$ we have%
\begin{equation}
\left\langle a_{l^{\prime}m^{\prime}}^{\ast}a_{lm}\right\rangle =\frac
{(-i)^{l^{\prime}}i^{l}}{2\pi^{2}}\sum_{i=1}^{3}I_{i}(l^{\prime},m^{\prime
};l,m)F_{i}(l^{\prime},l)\ . \label{offdiag}%
\end{equation}
Note that the off-diagonal terms depend on our anisotropic functions
$f_{i}(x)$ (given by (\ref{fns})), where these were determined by our
isotropic deficit function $\xi(k)$ via the equations (\ref{fns_defn}). Thus
the off-diagonal terms contain information about the form of the function
$\xi(k)$, which our simulations have constrained to be (approximately) an inverse-tangent.

Considering only the lowest ($L=1$) terms in the expansions (\ref{delta_expn2}%
) for $\Delta_{i}(\mathbf{\hat{k}})$, we may study the integrals
$I_{i}(l^{\prime},m^{\prime};l,m)$ and obtain the lowest-order anisotropic
(off-diagonal) corrections (\ref{offdiag}) to the covariance
matrix.\footnote{In future work we hope to consider these effects in terms of
the convenient formalism of bipolar spherical harmonics \cite{HS05,Kumar14}.
The approach taken here suffices for our present purposes.}

\subsection{Multipole $l-(l+1)$ correlations}

We first evaluate the integrals (\ref{I_i}), with $\mathbf{\hat{k}}$
represented by angular coordinates $(\theta,\phi)$. Considering only the
lowest ($L=1$) terms in the expansions (\ref{delta_expn2}) for $\Delta
_{i}(\mathbf{\hat{k}})$, we have%
\begin{align}
I_{i}(l^{\prime},m^{\prime};l,m) &  =\Delta_{10}^{i}\sqrt{3/4\pi}\int
d\Omega\ Y_{l^{\prime}m^{\prime}}^{\ast}(\theta,\phi)\cos\theta.Y_{lm}%
(\theta,\phi)\nonumber\\
&  -\Delta_{11}^{i}\sqrt{3/8\pi}\int d\Omega\ Y_{l^{\prime}m^{\prime}}^{\ast
}(\theta,\phi)e^{i\phi}\sin\theta.Y_{lm}(\theta,\phi)\label{Iseries}\\
&  -(\Delta_{11}^{i})^{\ast}\sqrt{3/8\pi}\int d\Omega\ Y_{l^{\prime}m^{\prime
}}^{\ast}(\theta,\phi)e^{-i\phi}\sin\theta.Y_{lm}(\theta,\phi)\ .\nonumber
\end{align}
We may use the following relations (ref. \cite{Arf}, p. 805):%
\begin{align}
\cos\theta.Y_{lm}(\theta,\phi) &  =\alpha_{1}Y_{(l+1)m}(\theta,\phi
)+\alpha_{2}Y_{(l-1)m}(\theta,\phi)\ ,\nonumber\\
e^{i\phi}\sin\theta.Y_{lm}(\theta,\phi) &  =-\alpha_{3}Y_{(l+1)(m+1)}%
(\theta,\phi)+\alpha_{4}Y_{(l-1)(m+1)}(\theta,\phi)\ ,\label{Arfken}\\
e^{-i\phi}\sin\theta.Y_{lm}(\theta,\phi) &  =\alpha_{5}Y_{(l+1)(m-1)}%
(\theta,\phi)-\alpha_{6}Y_{(l-1)(m-1)}(\theta,\phi)\ ,\nonumber
\end{align}
where%
\begin{align}
\alpha_{1} &  =\sqrt{\frac{(l-m+1)(l+m+1)}{(2l+1)(2l+3)}}\ ,\ \ \ \alpha
_{2}=\sqrt{\frac{(l-m)(l+m)}{(2l-1)(2l+1)}}\ ,\nonumber\\
\alpha_{3} &  =\sqrt{\frac{(l+m+1)(l+m+2)}{(2l+1)(2l+3)}}\ ,\ \ \ \alpha
_{4}=\sqrt{\frac{(l-m)(l-m-1)}{(2l-1)(2l+1)}}\ ,\label{coeffs}\\
\alpha_{5} &  =\sqrt{\frac{(l-m+1)(l-m+2)}{(2l+1)(2l+3)}}\ ,\ \ \ \alpha
_{6}=\sqrt{\frac{(l+m)(l+m-1)}{(2l-1)(2l+1)}}\ .\nonumber
\end{align}
The integrals in (\ref{Iseries}) then follow trivially from the orthonormality
relations $\int d\Omega\ Y_{l^{\prime}m^{\prime}}^{\ast}Y_{lm}=\delta
_{ll^{\prime}}\delta_{mm^{\prime}}$. We find%
\begin{align}
I_{i}(l^{\prime},m^{\prime};l,m) &  =\delta_{l^{\prime}(l+1)}\sqrt{3/4\pi
}\left[  \Delta_{10}^{i}\alpha_{1}\delta_{m^{\prime}m}+\Delta_{11}^{i}\frac
{1}{\sqrt{2}}\alpha_{3}\delta_{m^{\prime}(m+1)}-(\Delta_{11}^{i})^{\ast}%
\frac{1}{\sqrt{2}}\alpha_{5}\delta_{m^{\prime}(m-1)}\right]  \nonumber\\
&  +\delta_{l^{\prime}(l-1)}\sqrt{3/4\pi}\left[  \Delta_{10}^{i}\alpha
_{2}\delta_{m^{\prime}m}-\Delta_{11}^{i}\frac{1}{\sqrt{2}}\alpha_{4}%
\delta_{m^{\prime}(m+1)}+(\Delta_{11}^{i})^{\ast}\frac{1}{\sqrt{2}}\alpha
_{6}\delta_{m^{\prime}(m-1)}\right]  \ .\label{Iresult}%
\end{align}

Note that%
\[
I_{i}(l,m;l,m)=0\ .
\]
Thus for $l^{\prime}m^{\prime}=lm$ we have simply%
\begin{equation}
\left\langle \left\vert a_{lm}\right\vert ^{2}\right\rangle =C_{l}\ ,
\label{diag}%
\end{equation}
where $C_{l}$ is the corrected angular power spectrum (\ref{C_l_corr}) with
deficit function $\xi(k)$.

From (\ref{Iresult}) for $l^{\prime}m^{\prime}\neq lm$, the only non-zero
off-diagonal values of $I_{i}(l^{\prime},m^{\prime};l,m)$ are%
\begin{equation}
I_{i}(l+1,m^{\prime};l,m)=\sqrt{3/4\pi}\left[  \Delta_{10}^{i}\alpha_{1}%
\delta_{m^{\prime}m}+\Delta_{11}^{i}\frac{1}{\sqrt{2}}\alpha_{3}%
\delta_{m^{\prime}(m+1)}-(\Delta_{11}^{i})^{\ast}\frac{1}{\sqrt{2}}\alpha
_{5}\delta_{m^{\prime}(m-1)}\right]  \label{I-l+1}%
\end{equation}
and%
\begin{equation}
I_{i}(l-1,m^{\prime};l,m)=\sqrt{3/4\pi}\left[  \Delta_{10}^{i}\alpha_{2}%
\delta_{m^{\prime}m}-\Delta_{11}^{i}\frac{1}{\sqrt{2}}\alpha_{4}%
\delta_{m^{\prime}(m+1)}+(\Delta_{11}^{i})^{\ast}\frac{1}{\sqrt{2}}\alpha
_{6}\delta_{m^{\prime}(m-1)}\right]  \ .\label{I-l-1}%
\end{equation}

Thus, from (\ref{offdiag}), the only non-vanishing off-diagonal correlations
$\left\langle a_{l^{\prime}m^{\prime}}^{\ast}a_{lm}\right\rangle $ are%
\begin{equation}
\left\langle a_{(l+1)m^{\prime}}^{\ast}a_{lm}\right\rangle =-\frac{i}{2\pi
^{2}}\sum_{i=1}^{3}I_{i}(l+1,m^{\prime};l,m)F_{i}(l+1,l) \label{offdiag1}%
\end{equation}
and%
\begin{equation}
\left\langle a_{(l-1)m^{\prime}}^{\ast}a_{lm}\right\rangle =\frac{i}{2\pi^{2}%
}\sum_{i=1}^{3}I_{i}(l-1,m^{\prime};l,m)F_{i}(l-1,l)\ , \label{offdiag2}%
\end{equation}
where%
\begin{equation}
F_{i}(l\pm1,l)=\int_{0}^{\infty}\frac{dk}{k}\ \mathcal{T}(k,l\pm
1)\mathcal{T}(k,l)\mathcal{P}_{\mathcal{R}}^{\mathrm{QT}}(k)f_{i}(x)\ .
\label{F_i_l+-1}%
\end{equation}

We emphasise that the off-diagonal terms (\ref{offdiag1}) and (\ref{offdiag2})
in the covariance matrix depend on the form of our isotropic deficit function
$\xi(k)$ (via the associated functions $f_{i}(x)$).

\section{Anisotropic signatures of cosmological quantum relaxation}

The aim of this paper is to find signatures of cosmological quantum relaxation
in statistical CMB anisotropies. In particular, it would be desirable to have
signatures that are independent of the unknown -- and as far as we know
arbitrary -- coefficients $\Delta_{10}^{i}$ and $\Delta_{11}^{i}$ appearing in
the anisotropic expansions (\ref{delta_expn2}). Such signatures will be
presented in this section. Whether or not they might be detectable in practice
is outside the scope of this paper, but some relevant comments will be made in
the concluding Section 6.

As we have noted, the quantum-theoretical spectrum $\mathcal{P}_{\mathcal{R}%
}^{\mathrm{QT}}(k)$ is necessarily isotropic because the quantum equilibrium
width of inflationary vacuum modes is independent of the direction
$\mathbf{\hat{k}}$ of the wave vector $\mathbf{k}$. On the other hand, early
quantum nonequilibrium is expected to be anisotropic: the nonequilibrium width
of the relevant field modes can generally depend on $\mathbf{\hat{k}}$. In our
quantum relaxation scenario, we expect to find residual nonequilibrium at long
wavelengths and (at least)\ approximate equilibrium at short wavelengths --
where the transition from one regime to another is precisely characterised by
our inverse-tangent deficit function $\xi(k)$ (equation (\ref{numksi}))
obtained from numerical simulations. As the wavenumber $k$ increases,
equilibrium is approached and the anisotropic nonequilibrium width approaches
(at least approximately) the isotropic equilibrium width. Thus anisotropy at
large scales necessarily gives way to approximate isotropy at small scales.

Now because we have a precise characterisation of how the width deficit
$\xi(k)$ varies with $k$, in the limit of weak anisotropy we are able to
deduce how the anisotropies in the generalised deficit $\xi(k,\mathbf{\hat{k}%
})$ vary with $k$ (via the argument given in Section 3.3). As a result, we are
able to make predictions for the behaviour of the anisotropies in the
covariance matrix as a function of multipole moment $l$. Furthermore, we are
able to predict residual statistical anisotropies at small scales.

These results amount to constraints, or consistency relations, between the
impact of quantum nonequilibrium on the power deficit and the impact of
quantum nonequilibrium on statistical anisotropy. In our scenario, the two
kinds of anomaly are closely related and we are in fact able to deduce
quantitative relationships between them. In principle, the deduced
relationships could be tested against CMB data (though we do not attempt to do
so here).

Our scenario is also likely to have implications for the question of anomalous
mode alignment at large scales. This is discussed briefly in Section 5.4.

\subsection{Low-$l$ scaling of $l-(l+1)$ correlations with $l$}

Let us study how our correlation matrix elements $\left\langle
a_{(l+1)m^{\prime}}^{\ast}a_{lm}\right\rangle $ -- as given by (\ref{offdiag1}%
) -- scale as a function of multipole moment $l$ in the low-$l$ regime. (The
other non-zero elements $\left\langle a_{(l-1)m^{\prime}}^{\ast}%
a_{lm}\right\rangle $ are trivially equal to $\left\langle a_{lm}^{\ast
}a_{(l-1)m^{\prime}}\right\rangle ^{\ast}$.)

We wish to extract an approximate scaling with $l$ that is independent of the
unknown coefficients $\Delta_{10}^{i}$ and $\Delta_{11}^{i}$ in the expansions
(\ref{delta_expn2}). To this end, we consider the average over $m^{\prime}$
and $m$:%
\begin{equation}
\overline{\left\langle a_{(l+1)m^{\prime}}^{\ast}a_{lm}\right\rangle }%
\equiv\frac{1}{2(l+1)+1}\frac{1}{2l+1}\sum_{m^{\prime}=-(l+1)}^{(l+1)}%
\sum_{m=-l}^{l}\left\langle a_{(l+1)m^{\prime}}^{\ast}a_{lm}\right\rangle \ .
\end{equation}
From (\ref{offdiag1}) this may be written as%
\begin{equation}
\overline{\left\langle a_{(l+1)m^{\prime}}^{\ast}a_{lm}\right\rangle }%
=-\frac{i}{2\pi^{2}}\sum_{i=1}^{3}\overline{I_{i}(l+1,m^{\prime};l,m)}%
F_{i}(l+1,l)\ , \label{avcovmat}%
\end{equation}
where%
\begin{equation}
\overline{I_{i}(l+1,m^{\prime};l,m)}\equiv\frac{1}{2(l+1)+1}\frac{1}{2l+1}%
\sum_{m^{\prime}=-(l+1)}^{(l+1)}\sum_{m=-l}^{l}I_{i}(l+1,m^{\prime};l,m)\ .
\label{avI}%
\end{equation}

Let us consider the integral factors in (\ref{avcovmat}). As a simple
approximation we may take $\mathcal{P}_{\mathcal{R}}^{\mathrm{QT}}(k)\simeq
const.\equiv\mathcal{P}_{\mathcal{R}}^{\mathrm{QT}}$. At low $l$ the transfer
function takes the form \cite{LL00}%
\begin{equation}
\mathcal{T}(k,l)=\sqrt{\pi}H_{0}^{2}\ j_{l}(2k/H_{0}) \label{transfn}%
\end{equation}
(where $H_{0}$ is the Hubble parameter today). We take (\ref{transfn}) to be
valid for (say) $l\lesssim20$ -- ignoring the small contribution from the
integrated Sachs-Wolfe effect at very low $l$. In this range of $l$ we then
have%
\begin{equation}
F_{i}(l+1,l)\simeq\pi H_{0}^{4}\mathcal{P}_{\mathcal{R}}^{\mathrm{QT}}h_{i}(l)
\label{int_l+1_l}%
\end{equation}
with%
\begin{equation}
h_{i}(l)\equiv\int_{0}^{\infty}\frac{dx}{x}\ j_{l+1}(x)j_{l}(x)f_{i}(x)\ ,
\label{h(l)}%
\end{equation}
where again it is convenient to work with the parameter $x=2k/H_{0}$.

Our averaged covariance matrix elements (\ref{avcovmat}) then take the form%
\begin{equation}
\overline{\left\langle a_{(l+1)m^{\prime}}^{\ast}a_{lm}\right\rangle }%
=-\frac{i}{2\pi}H_{0}^{4}\mathcal{P}_{\mathcal{R}}^{\mathrm{QT}}\sum_{i=1}%
^{3}\overline{I_{i}(l+1,m^{\prime};l,m)}h_{i}(l)\ . \label{avcovmat2}%
\end{equation}
The dependence on $l$ is contained in the factors $\overline{I_{i}%
(l+1,m^{\prime};l,m)}$ and $h_{i}(l)$. Let us study these in turn.

From (\ref{I-l+1}) we have%
\begin{equation}
\overline{I_{i}(l+1,m^{\prime};l,m)}=\frac{1}{2}\sqrt{\frac{3}{\pi}}\frac
{1}{2l+3}\frac{1}{2l+1}\left(  \Delta_{10}^{i}s_{1}+\Delta_{11}^{i}\frac
{1}{\sqrt{2}}s_{3}-(\Delta_{11}^{i})^{\ast}\frac{1}{\sqrt{2}}s_{5}\right)
\label{Ibar}%
\end{equation}
where
\[
s_{1}\equiv\sum_{m=-l}^{l}\alpha_{1}(l,m)\ ,\ s_{3}\equiv\sum_{m=-l}^{l}%
\alpha_{3}(l,m)\ ,\ s_{5}\equiv\sum_{m=-l}^{l}\alpha_{5}(l,m)\ .
\]
Writing%
\[
s_{n}=\frac{1}{\sqrt{(2l+1)(2l+3)}}\sigma_{n}(l)
\]
(for $n=1,$ $3$, $5$) and using (\ref{coeffs}), in the region $l=2$ to $l=20$
we find by numerical evaluation that%
\begin{align*}
\sigma_{1}(l)  &  =\sum_{m=-l}^{l}\sqrt{(l-m+1)(l+m+1)}\approx2.51\times
l^{15/8}\ ,\\
\sigma_{3}(l)  &  =\sum_{m=-l}^{l}\sqrt{(l+m+1)(l+m+2)}\approx3.22\times
l^{15/8}\ .
\end{align*}
Also,%
\[
\sigma_{5}(l)=\sum_{m=-l}^{l}\sqrt{(l-m+1)(l-m+2)}=\sum_{m=-l}^{l}%
\sqrt{(l+m+1)(l+m+2)}=\sigma_{3}(l)\ .
\]
For all three terms we have $\sigma_{n}(l)\propto l^{15/8}$ (to a good
approximation). If we take $1/\sqrt{(2l+1)(2l+3)}\sim1/l$ -- which is fairly
accurate except for the very lowest $l$ -- we find the same approximate
scaling%
\begin{equation}
s_{n}\propto l^{7/8}%
\end{equation}
for all three terms appearing in (\ref{Ibar}). Taking $\frac{1}{2l+3}\frac
{1}{2l+1}\sim1/l^{2}$ -- again fairly accurate except for the very lowest $l$
-- we then have $\overline{I_{i}(l+1,m^{\prime};l,m)}\sim l^{-9/8}$ or simply%
\begin{equation}
\overline{I_{i}(l+1,m^{\prime};l,m)}\sim l^{-1} \label{Ibarscaling}%
\end{equation}
(where $l^{-9/8}$ agrees closely with $0.85l^{-1}$ in this range of $l$).

Let us now study the factors $h_{i}(l)$. Taking the illustrative values
(\ref{fidc's}) for $\tilde{c}_{1}$, $c_{2}$, we have%
\begin{equation}
f_{1}(x)=\frac{x}{1+(0.5x+3)^{2}}\ ,\ \ \ \ f_{2}(x)=\frac{1}{1+(0.5x+3)^{2}%
}\ ,\ \ \ \ f_{3}(x)=1\ ,\label{fns_fid}%
\end{equation}
and%
\begin{equation}
h_{1}(l)=\int_{0}^{\infty}\frac{dx}{x}\ j_{l+1}(x)j_{l}(x)\frac{x}%
{1+(0.5x+3)^{2}}\ ,\label{h(l)_1}%
\end{equation}%
\begin{equation}
h_{2}(l)=\int_{0}^{\infty}\frac{dx}{x}\ j_{l+1}(x)j_{l}(x)\frac{1}%
{1+(0.5x+3)^{2}}\ ,\label{h(l)_2}%
\end{equation}%
\begin{equation}
h_{3}(l)=\int_{0}^{\infty}\frac{dx}{x}\ j_{l+1}(x)j_{l}(x)\ .\label{h(l)_3}%
\end{equation}
The integrals (\ref{h(l)_1}), (\ref{h(l)_2}), (\ref{h(l)_3}) may be evaluated
numerically.\footnote{I am grateful to Samuel Colin for assistance with the
numerical evaluation of these integrals.} Results for $l=2$, $3$, $4$, ... ,
$20$ are displayed in Figure 2.\textbf{ }We find close fits to the curves%
\begin{equation}
h_{1}(l)=\allowbreak0.018\times l^{-3/2}\ ,\ \ \ \ h_{2}(l)=0.0065\times
l^{-2}\ ,\ \ \ \ h_{3}(l)=0.12\times l^{-3/2}\ .\label{h_scaling}%
\end{equation}
We may conclude that $h_{1},h_{3}\propto l^{-3/2}$ and $h_{2}\propto l^{-2}$
to a good approximation (in this low-$l$ region). These scalings are peculiar
to our inverse-tangent deficit function (\ref{numksi}) (cf. Section 5.2).%

\begin{figure}
[ptb]
\begin{center}
\includegraphics[width=0.7\textwidth]{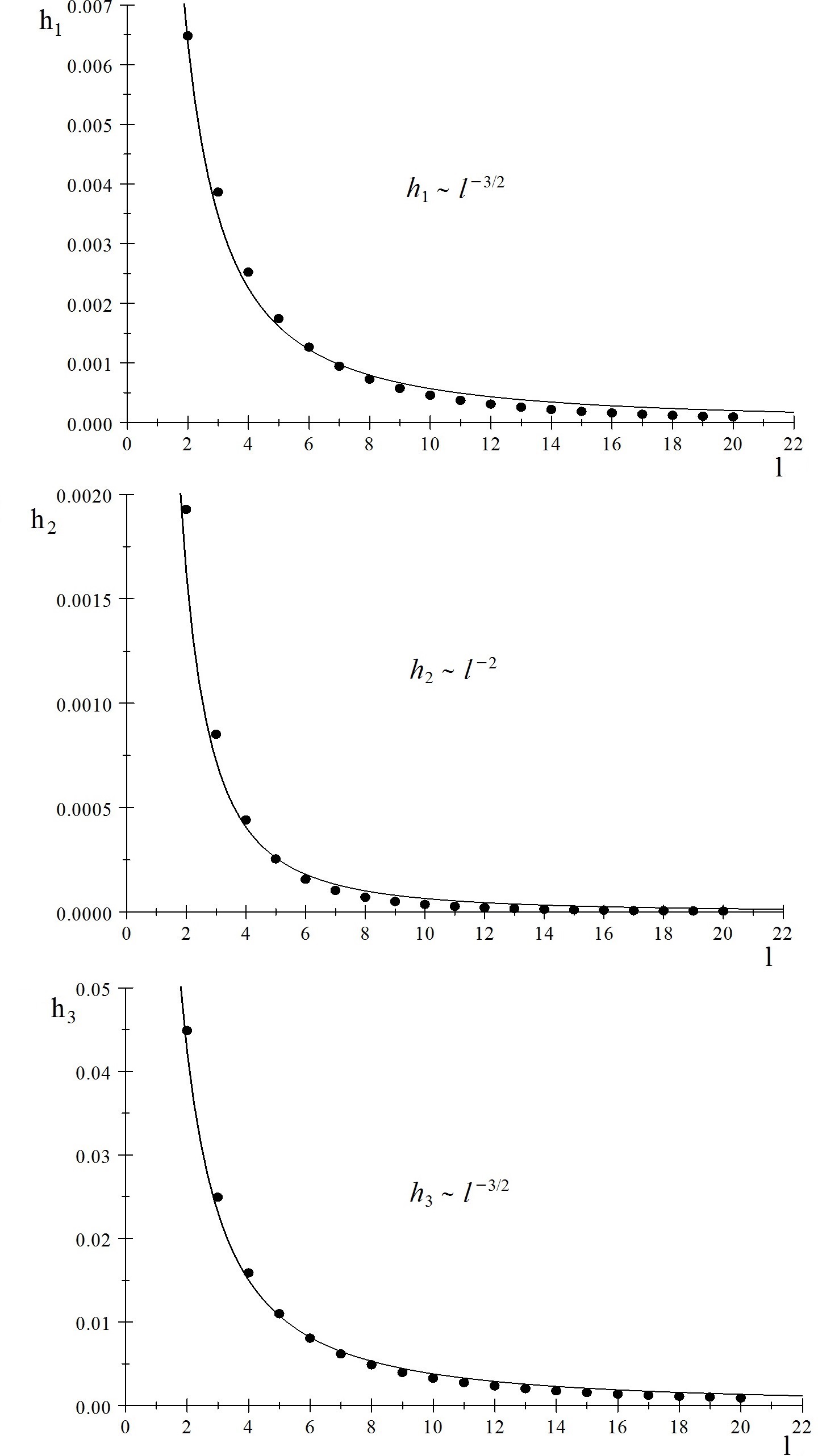}
\caption{Numerical results for the integrals (\ref{h(l)_1}), (\ref{h(l)_2}),
(\ref{h(l)_3}) in the range $l=2-20$, with fits to the respective curves
$h_{1}\propto l^{-3/2}$, $h_{2}\propto l^{-2}$, $h_{3}\propto l^{-3/2}$.}%
\end{center}
\end{figure}

Putting our results (\ref{Ibarscaling}) and (\ref{h_scaling}) together, we
find that different terms in (\ref{avcovmat2}) can scale with two different
powers of $l$. For $i=1,3$ we have terms that scale as $\sim l^{-1}\times
l^{-3/2}=l^{-5/2}$, while for $i=2$ we have terms that scale as $\sim
l^{-1}\times l^{-2}=l^{-3}$. Thus the terms in $\overline{\left\langle
a_{(l+1)m^{\prime}}^{\ast}a_{lm}\right\rangle }$ will scale as%
\begin{equation}
\overline{\left\langle a_{(l+1)m^{\prime}}^{\ast}a_{lm}\right\rangle }\sim
l^{-5/2},\ l^{-3}\ . \label{avcovmatscalings}%
\end{equation}

We conclude that, for our inverse-tangent deficit function (\ref{numksi}), the
averaged $l-(l+1)$ covariance matrix elements (\ref{avcovmat2}) will contain
terms that scale approximately as $l^{-5/2}$ or as $l^{-3}$ at low $l$ --
regardless of the values of $\Delta_{10}^{i}$, $\Delta_{11}^{i}$ in the
expansions (\ref{delta_expn2}).

\subsection{Low-$l$ correlation scaling with an alternative deficit function}

The scaling of $\overline{\left\langle a_{(l+1)m^{\prime}}^{\ast}%
a_{lm}\right\rangle }$ with $l$ depends on the form of the deficit function
$\xi(k)$. We may illustrate this by considering an alternative deficit function.

We take a three-parameter function $\xi_{\mathrm{dip}}(k)$ with
near-equilibrium ($\xi_{\mathrm{dip}}=\kappa_{3}$ where $\kappa_{3}$ is
understood to be close to $1$) for $x>\kappa_{2}$ and a sudden dip
($\xi_{\mathrm{dip}}=\kappa_{1}<\kappa_{3}$) in power for $x<\kappa_{2}$:%
\begin{align*}
\xi_{\mathrm{dip}}(k) &  =\kappa_{1}\ \ \ \mathrm{for}\ x<\kappa_{2}\ ,\\
\xi_{\mathrm{dip}}(k) &  =\kappa_{3}\ \ \ \mathrm{for}\ x>\kappa_{2}\ .
\end{align*}
We may write this as%
\begin{equation}
\xi_{\mathrm{dip}}(k)=\kappa_{1}\theta(\kappa_{2}-x)+\kappa_{3}\theta
(x-\kappa_{2})\ .\label{ksi_alt}%
\end{equation}

Let us follow the same procedure as before, generalising the coefficients
$\kappa_{i}$ ($i=1,2,3$) to direction-dependent functions $\kappa
_{i}(\mathbf{\hat{k}})$, which we write as $\kappa_{i}(\mathbf{\hat{k}%
})=\kappa_{i}+\Delta_{i}(\mathbf{\hat{k}})$ with $\Delta_{i}(\mathbf{\hat{k}%
})$ assumed small and expanded as in (\ref{expns'}). Our alternative isotropic
function (\ref{ksi_alt}) is then generalised to the alternative anisotropic
function%
\begin{equation}
\xi_{\mathrm{dip}}(k,\mathbf{\hat{k}})=\left(  \kappa_{1}+\Delta
_{1}(\mathbf{\hat{k}})\right)  \theta(\kappa_{2}+\Delta_{2}(\mathbf{\hat{k}%
})-x)+\left(  \kappa_{3}+\Delta_{3}(\mathbf{\hat{k}})\right)  \theta
(x-\kappa_{2}-\Delta_{2}(\mathbf{\hat{k}}))\ .
\end{equation}
Expanding to lowest order in $\Delta_{i}(\mathbf{\hat{k}})$ and defining the
alternative functions%
\begin{equation}
f_{1}^{\mathrm{dip}}(x)\equiv\frac{\partial\xi_{\mathrm{dip}}(k)}%
{\partial\kappa_{1}}\ ,\ \ \ f_{2}^{\mathrm{dip}}(x)\equiv\frac{\partial
\xi_{\mathrm{dip}}(k)}{\partial\kappa_{2}}\ ,\ \ \ f_{3}^{\mathrm{dip}%
}(x)\equiv\frac{\partial\xi_{\mathrm{dip}}(k)}{\partial\kappa_{3}}\ ,
\end{equation}
we have%
\begin{equation}
\xi_{\mathrm{dip}}(k,\mathbf{\hat{k}})=\xi_{\mathrm{dip}}(k)+\sum_{i=1}%
^{3}f_{i}^{\mathrm{dip}}(x)\Delta_{i}(\mathbf{\hat{k}})\label{ksi-dip_aniso}%
\end{equation}
where%
\begin{equation}
f_{1}^{\mathrm{dip}}(x)=\theta(\kappa_{2}-x)\ ,\ \ \ \ f_{2}^{\mathrm{dip}%
}(x)=(\kappa_{1}-\kappa_{3})\delta(x-\kappa_{2})\ ,\ \ \ \ f_{3}%
^{\mathrm{dip}}(x)=\theta(x-\kappa_{2})\ .
\end{equation}
In this case the anisotropic part of $\xi_{\mathrm{dip}}(k,\mathbf{\hat{k}})$
depends on all three parameters $\kappa_{1}$, $\kappa_{2}$, $\kappa_{3}$.

Note that, despite the appearance of the delta-function $\delta(x-\kappa_{2})$
in $f_{2}^{\mathrm{dip}}(x)$, for small $\Delta_{2}(\mathbf{\hat{k}})$ the
right-hand side of (\ref{ksi-dip_aniso}) is a good approximation to the
left-hand side (to first order in $\Delta_{2}(\mathbf{\hat{k}})$) when both
sides appear under an integral $\int dx$ with a function that varies little
over a distance $\Delta_{2}(\mathbf{\hat{k}})$. In our case, where
$\xi_{\mathrm{dip}}(k,\mathbf{\hat{k}})$ appears under the integral
(\ref{covmat2}) for $\left\langle a_{l^{\prime}m^{\prime}}^{\ast}%
a_{lm}\right\rangle $, at low $l$ we require that $\Delta_{2}(\mathbf{\hat{k}%
})$ be small compared with the scale over which $j_{l+1}(x)j_{l}(x)$ varies
with $x$.

Let us consider a specific alternative function (\ref{ksi_alt}) that is
numerically not far from the inverse-tangent function (\ref{numksi}) with the
illustrative values (\ref{fidc's}) for $\tilde{c}_{1}$, $c_{2}$ and (for
simplicity) with $c_{3}=1$. We take%
\begin{align}
\xi_{\mathrm{dip}}(k) &  =0.7\ \ \ \mathrm{for}\ x<20\ ,\nonumber\\
\xi_{\mathrm{dip}}(k) &  =1\ \ \ \mathrm{for}\ x>20\ .\label{ksi-dip-fid}%
\end{align}
(The area under the curve $1-\xi_{\mathrm{dip}}(k)$ is then equal to $6$,
which approximately matches the value of $\int_{0}^{\infty}(1-\xi(k))dx$. See
Figure 3.) This corresponds to a deficit function $\xi_{\mathrm{dip}}(k)$ with
equilibrium ($\xi_{\mathrm{dip}}=1$) for $x>20$ and a sudden dip
($\xi_{\mathrm{dip}}=0.7$) in power for $x<20$ -- that is, a dip for
$\lambda>(\pi/5)H_{0}^{-1}$. In terms of our parameters, we have%
\begin{equation}
\kappa_{1}=0.7\ ,\ \ \ \kappa_{2}=20\ ,\ \ \ \kappa_{3}=1\ .
\end{equation}
Our alternative `fiducial' functions are then%
\begin{equation}
f_{1}^{\mathrm{dip}}(x)=\theta(20-x)\ ,\ \ \ \ f_{2}^{\mathrm{dip}%
}(x)=-0.3\delta(x-20)\ ,\ \ \ \ f_{3}^{\mathrm{dip}}(x)=\theta(x-20)\ .
\end{equation}
%

\begin{figure}
[ptb]
\begin{center}
\includegraphics[width=0.7\textwidth]{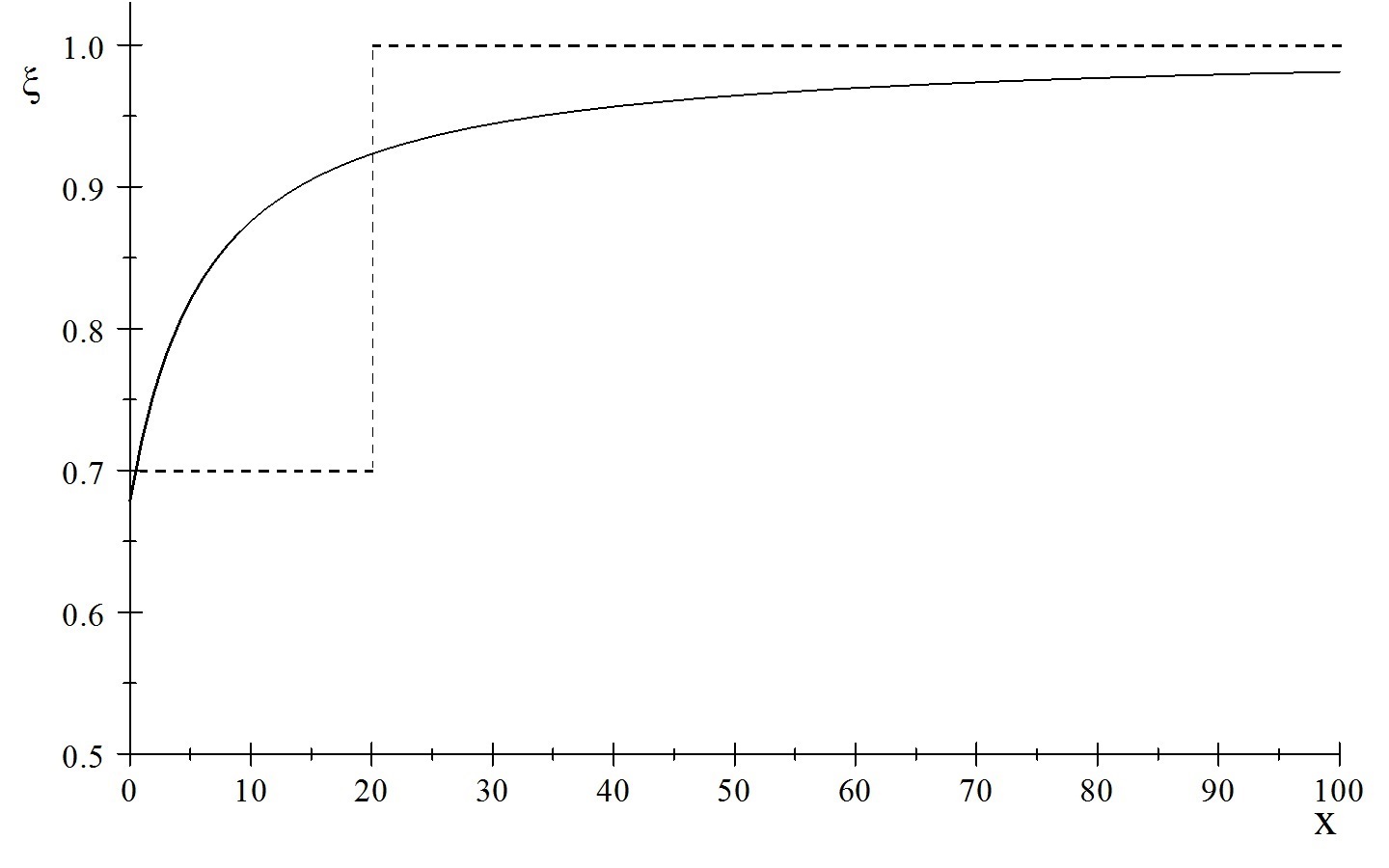}
\caption{The deficit function $\xi=\tan^{-1}(0.5x+3)-\frac{\pi}{2}+1$ (solid
line, with $x=2k/H_{0}$) arising from cosmological quantum relaxation and the
alternative function $\xi_{\mathrm{dip}}$ (dashed line, given by
(\ref{ksi-dip-fid})) corresponding to a sudden dip in power.}%
\end{center}
\end{figure}

We then have the alternative integrals:%
\begin{equation}
h_{1}^{\mathrm{dip}}(l)=\int_{0}^{\infty}\frac{dx}{x}\ j_{l+1}(x)j_{l}%
(x)\theta(20-x)=\int_{0}^{20}\frac{dx}{x}\ j_{l+1}(x)j_{l}%
(x)\ ,\label{h(l)_dip_1}%
\end{equation}%
\begin{equation}
h_{2}^{\mathrm{dip}}(l)=\int_{0}^{\infty}\frac{dx}{x}\ j_{l+1}(x)j_{l}%
(x)\left[  -0.3\delta(x-20)\right]  =-\frac{3}{200}j_{l+1}(20)j_{l}%
(20)\ ,\label{h(l)_dip_2}%
\end{equation}%
\begin{equation}
h_{3}^{\mathrm{dip}}(l)=\int_{0}^{\infty}\frac{dx}{x}\ j_{l+1}(x)j_{l}%
(x)\theta(x-20)=\int_{20}^{\infty}\frac{dx}{x}\ j_{l+1}(x)j_{l}%
(x)\ .\label{h(l)_dip_3}%
\end{equation}
These may again be evaluated numerically. Results for $l=2$, $3$, $4$, ... ,
$20$ are displayed in Figure 4. For $h_{1}^{\mathrm{dip}}(l)$ and
$h_{3}^{\mathrm{dip}}(l)$ we find (respectively close and approximate) fits to
the curves%
\begin{equation}
h_{1}^{\mathrm{dip}}(l)=0.12\times l^{-3/2}\ ,\ \ \ \ h_{3}^{\mathrm{dip}%
}(l)=7\times10^{-5}+1.85\times10^{-6}l^{2}\ ,\label{h-dip_scaling}%
\end{equation}
where in the latter case there are small oscillations around the fitted curve.
We may conclude that $h_{1}^{\mathrm{dip}}\sim l^{-3/2}$ and $h_{3}%
^{\mathrm{dip}}\sim\mathrm{const}.+l^{2}$ (in this low-$l$ region). For
$h_{2}^{\mathrm{dip}}(l)$, there are strong oscillations that grow with $l$.
These simple scalings for $h_{1}^{\mathrm{dip}}(l)$ and $h_{3}^{\mathrm{dip}%
}(l)$, together with the oscillatory behaviour of $h_{2}^{\mathrm{dip}}(l)$,
are peculiar to the alternative deficit function (\ref{ksi_alt}).%

\begin{figure}
[ptb]
\begin{center}
\includegraphics[width=0.7\textwidth]{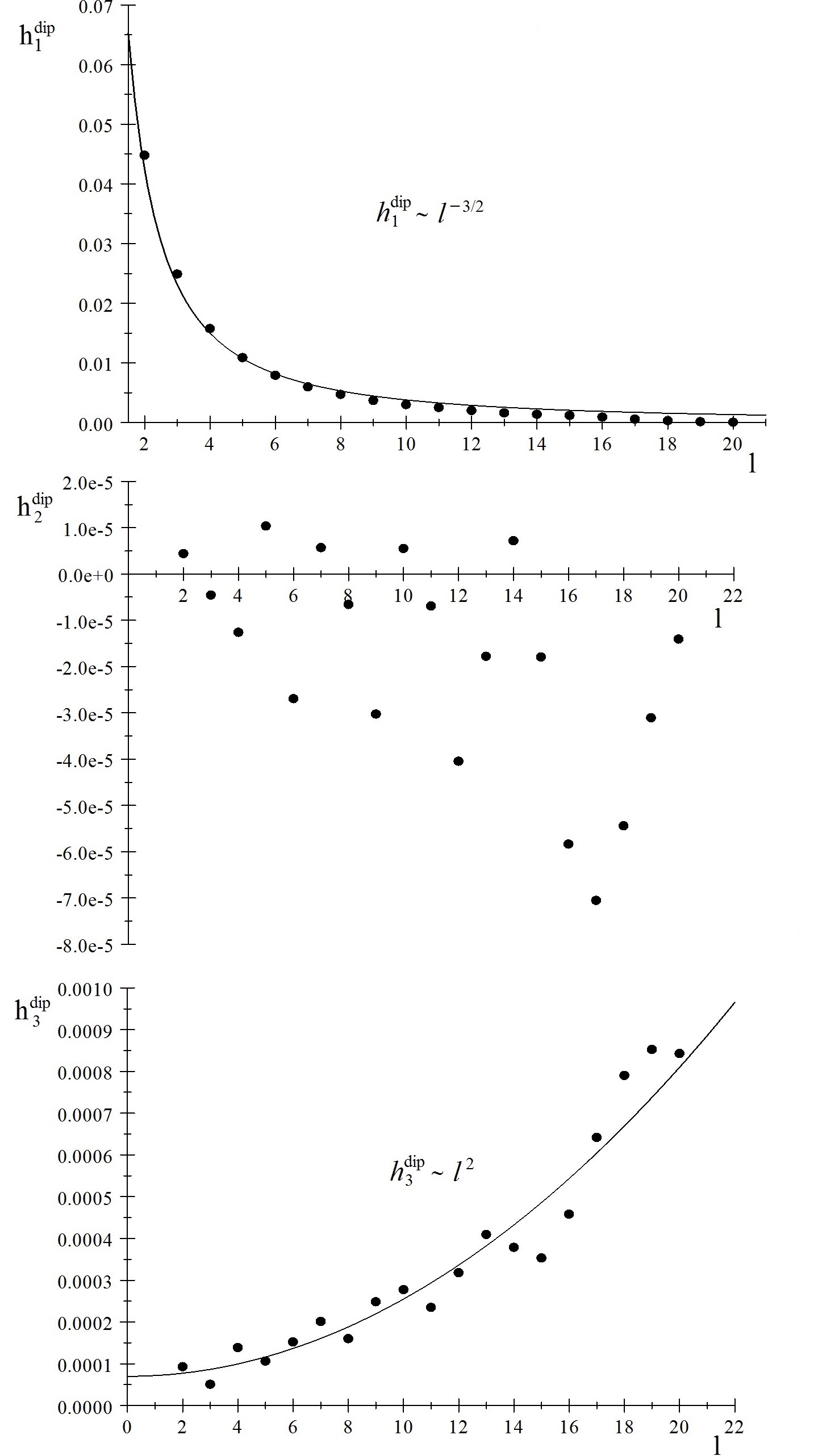}
\caption{Numerical results for the integrals (\ref{h(l)_dip_1}),
(\ref{h(l)_dip_2}), (\ref{h(l)_dip_3}) in the range $l=2-20$, with fitted
curves $h_{1}^{\mathrm{dip}}\sim l^{-3/2}$ and $h_{3}^{\mathrm{dip}}%
\sim\mathrm{const}.+l^{2}$.}%
\end{center}
\end{figure}

The averaged $l-(l+1)$ covariance matrix elements now take the form (from
(\ref{avcovmat2}))%
\[
\overline{\left\langle a_{(l+1)m^{\prime}}^{\ast}a_{lm}\right\rangle
}_{\mathrm{dip}}=-\frac{i}{2\pi}H_{0}^{4}\mathcal{P}_{\mathcal{R}%
}^{\mathrm{QT}}\sum_{i=1}^{3}\overline{I_{i}(l+1,m^{\prime};l,m)}%
h_{i}^{\mathrm{dip}}(l)\ ,
\]
where we still have $\overline{I_{i}(l+1,m^{\prime};l,m)}\sim l^{-1}$
(equation (\ref{Ibarscaling})). Thus for $i=1$ we have terms in $\overline
{\left\langle a_{(l+1)m^{\prime}}^{\ast}a_{lm}\right\rangle }_{\mathrm{dip}}$
that scale as $\sim l^{-1}\times l^{-3/2}=l^{-5/2}$ (just as in the case of an
inverse-tangent deficit). For $i=2$ we have terms that oscillate strongly with
$l$. Finally, for $i=3$ we have terms that scale as $\sim l^{-1}%
\times(\mathrm{const}.+l^{2})$; that is, we have terms that scale as $\sim
l^{-1}$ and $\sim l$. Thus there will be terms in $\overline{\left\langle
a_{(l+1)m^{\prime}}^{\ast}a_{lm}\right\rangle }_{\mathrm{dip}}$ that scale as%
\begin{equation}
\overline{\left\langle a_{(l+1)m^{\prime}}^{\ast}a_{lm}\right\rangle
}_{\mathrm{dip}}\sim l^{-5/2},\ l^{-1},\ l \label{avcovmatscalings-dip}%
\end{equation}
(as well as terms that oscillate strongly with $l$).

We note the contrast with the results in Section 5.1 for the inverse-tangent
deficit, where we found that $\overline{\left\langle a_{(l+1)m^{\prime}}%
^{\ast}a_{lm}\right\rangle }\sim l^{-5/2},\ l^{-3}$. The scaling $\sim l^{-3}$
distinguishes the inverse-tangent case from the dip, while the scalings $\sim
l^{-1},\ l$ distinguish the case of a dip from the inverse-tangent. If at low
$l$ the observed averaged matrix elements $\overline{\left\langle
a_{(l+1)m^{\prime}}^{\ast}a_{lm}\right\rangle }$ were found to contain terms
that scale as $\sim l^{-3}$, this would be inconsistent with a sudden dip in
primordial power but consistent with our inverse-tangent deficit. On the other
hand if the observations found terms that scale as $\sim l^{-1}$ or $\sim l$,
this would be inconsistent with our inverse-tangent deficit but consistent
with a sudden dip.

We emphasise that these results are independent of the values of the
coefficients $\Delta_{10}^{i}$, $\Delta_{11}^{i}$ in the expansions
(\ref{delta_expn2}). They reflect properties of the isotropic power deficit
function $\xi(k)$. The two examples studied here -- the inverse-tangent and
the sudden dip -- suffice to illustrate how the scaling of $\overline
{\left\langle a_{(l+1)m^{\prime}}^{\ast}a_{lm}\right\rangle }$ with $l$
depends on the form of $\xi(k)$. It would be of interest to consider other
deficit functions, such as power laws, and to study the scaling of
$\overline{\left\langle a_{(l+1)m^{\prime}}^{\ast}a_{lm}\right\rangle }$ with
$l$ for different cases. We leave this for future work.

\subsection{Residual anisotropies at high $l$}

Our simulations of cosmological quantum relaxation generally yield $c_{3}$
slightly unequal to $1$, so that there is a small nonequilibrium residue in
the large-$k$ limit. We then expect to find residual statistical anisotropy
even in the limit of high $l$. As we noted in Section 2.3, there is evidence
in the data for statistical anisotropy at high $l$, together with anomalous
correlations between directional asymmetries at large and small scales
\cite{Planck15-XVI-IsoStats}.

Let us then consider our anisotropic deficit function (\ref{ksi_aniso_0}) at
small scales. In the large-$k$ limit we have%
\[
\xi(k)\simeq c_{3}\ ,\ \ \ f_{1}(x)\simeq0\ ,\ \ \ f_{2}(x)\simeq0\ ,
\]
while for all $k$ we have $\ f_{3}(x)=1$ exactly. Thus for large $k$ we have
simply%
\begin{equation}
\xi(k,\mathbf{\hat{k}})\simeq c_{3}+\Delta_{3}(\mathbf{\hat{k}})=c_{3}%
+\sum_{L\geq1,M}\Delta_{LM}^{3}Y_{LM}(\mathbf{\hat{k}}) \label{ksi_large_k}%
\end{equation}
(with $\Delta_{LM}^{3}$ presumed small). In terms of the expansion
(\ref{ksiexpn}), our coefficients (\ref{g_00}) and (\ref{g_LM}) read%
\[
g_{00}(k)=-\sqrt{4\pi}\left(  1-c_{3}\right)
\]
and (for $L\geq1$)%
\[
g_{LM}(k)=\Delta_{LM}^{3}\ .
\]

The large-$k$ expression (\ref{ksi_large_k}) indeed implies that small
statistical anisotropies will still be present even at the smallest
lengthscales. In fact, at very small lengthscales the anisotropies are scale-independent.

Let us consider the covariance matrix $\left\langle a_{l^{\prime}m^{\prime}%
}^{\ast}a_{lm}\right\rangle $ at small scales. Using the full anisotropic
deficit function (\ref{ksi_aniso_0}), and again taking only the lowest terms
in the expansions (\ref{delta_expn2}) of $\Delta_{i}(\mathbf{\hat{k}})$, we
have the general results (\ref{offdiag1}), (\ref{offdiag2}) for the only
non-vanishing off-diagonal covariance matrix elements $\left\langle
a_{(l\pm1)m^{\prime}}^{\ast}a_{lm}\right\rangle $, with the factors
$I_{i}(l\pm1,m^{\prime};l,m)$ given by (\ref{I-l+1}), (\ref{I-l-1}) and the
factors $F_{i}(l\pm1,l)$ given by (\ref{F_i_l+-1}). To calculate $\left\langle
a_{(l\pm1)m^{\prime}}^{\ast}a_{lm}\right\rangle $ at high $l$, we would need
to use the transfer function $\mathcal{T}(k,l)$ at high $l$. We leave a
detailed analysis for future work. Here we simply note that, in the limit of
large $l$, if we write $\mathcal{T}(k,l+1)\mathcal{T}(k,l)\simeq
\mathcal{T}^{2}(k,l)$ then%
\begin{equation}
F_{i}(l+1,l)\simeq F_{i}(l,l)=\int_{0}^{\infty}\frac{dk}{k}\ \mathcal{T}%
^{2}(k,l)\mathcal{P}_{\mathcal{R}}^{\mathrm{QT}}(k)f_{i}(x) \label{F_i_ll}%
\end{equation}
and (\ref{offdiag1}) becomes%
\begin{equation}
\left\langle a_{(l+1)m^{\prime}}^{\ast}a_{lm}\right\rangle \simeq-\frac
{i}{2\pi^{2}}\sum_{i=1}^{3}I_{i}(l+1,m^{\prime};l,m)F_{i}(l,l)
\label{covmat_high_l}%
\end{equation}
(where again the other non-zero elements $\left\langle a_{(l-1)m^{\prime}%
}^{\ast}a_{lm}\right\rangle $ are trivially equal to $\left\langle
a_{lm}^{\ast}a_{(l-1)m^{\prime}}\right\rangle ^{\ast}$).

Let us consider the factors $F_{i}(l,l)$. For $i=3$ we have $f_{3}(x)=1$ and
so (using (\ref{Cl2'}))%
\[
F_{3}(l,l)=\int_{0}^{\infty}\frac{dk}{k}\ \mathcal{T}^{2}(k,l)\mathcal{P}%
_{\mathcal{R}}^{\mathrm{QT}}(k)=2\pi^{2}C_{l}^{\mathrm{QT}}\ .
\]
We may relate this to the corrected $C_{l}$ as given by (\ref{C_l_corr}). We
are interested in the behaviour at large $l$. If we assume that in this regime
the integral in (\ref{C_l_corr}) is dominated by the large-$k$ limit of the
integrand -- for which $\xi(k)\simeq c_{3}$ -- we obtain%
\[
C_{l}\simeq c_{3}C_{l}^{\mathrm{QT}}\ .
\]
We then have the simple result%
\[
F_{3}(l,l)\simeq(2\pi^{2}/c_{3})C_{l}\ .
\]
For the terms with $i=1,2$, if again we assume that for large $l$ the integral
in (\ref{F_i_ll}) is dominated by the large-$k$ limit of the integrand -- for
which now $f_{1}(x)\simeq0$ and $f_{2}(x)\simeq0$ -- we then have%
\[
F_{1}(l,l)\simeq0\ ,\ \ \ F_{2}(l,l)\simeq0\ .
\]
In this approximation, then, from (\ref{covmat_high_l}) we have simply (for
large $l$)%
\begin{equation}
\left\langle a_{(l+1)m^{\prime}}^{\ast}a_{lm}\right\rangle \simeq-\frac
{i}{2\pi^{2}}I_{3}(l+1,m^{\prime};l,m)(2\pi^{2}/c_{3})C_{l}\ .
\label{consisreln}%
\end{equation}

The approximate result (\ref{consisreln}) amounts to a consistency relation
between the off-diagonal covariance matrix elements $\left\langle
a_{(l+1)m^{\prime}}^{\ast}a_{lm}\right\rangle $ and the angular power spectrum
$C_{l}$ at high $l$. Of course the factor $I_{3}(l+1,m^{\prime};l,m)$ contains
the unknown anisotropy coefficients $\Delta_{10}^{3}$ and $\Delta_{11}^{3}$,
but even so (\ref{consisreln}) relates the scaling of $\left\langle
a_{(l+1)m^{\prime}}^{\ast}a_{lm}\right\rangle $ with $l$ to the scaling of
$C_{l}$ with $l$.

Let us consider the factor $I_{3}(l+1,m^{\prime};l,m)$ for large $l$. From the
expressions (\ref{coeffs}), we find (for $l>>1$)%
\[
\alpha_{1}\simeq\frac{1}{2}\sqrt{1-m^{2}/l^{2}},\ \ \ \ \ \alpha_{3}%
\simeq\frac{1}{2}(1+m/l),\ \ \ \ \ \alpha_{5}\simeq\frac{1}{2}(1-m/l)\ .
\]
From (\ref{I-l+1}) we then have%
\begin{align}
I_{3}(l+1,m^{\prime};l,m)  & \simeq\frac{1}{2}\sqrt{3/4\pi}[\Delta_{10}%
^{3}\sqrt{1-m^{2}/l^{2}}\delta_{m^{\prime}m}+\\
& \Delta_{11}^{3}(1/\sqrt{2})(1+m/l)\delta_{m^{\prime}(m+1)}-(\Delta_{11}%
^{3})^{\ast}(1/\sqrt{2})(1-m/l)\delta_{m^{\prime}(m-1)}]\ .\nonumber
\end{align}
The dependence on $l$ is negligible for $m<<l$.

We may also consider the average (\ref{avI}) of $I_{3}(l+1,m^{\prime};l,m)$
over $m$ and $m^{\prime}$ (for large $l$). We find%
\begin{equation}
\overline{I_{3}(l+1,m^{\prime};l,m)}\simeq\frac{1}{16}\sqrt{\frac{3}{2\pi}%
}\frac{1}{l^{2}}\left[  \Delta_{10}^{3}\sqrt{2}\left(  1+1.58l\right)
+\Delta_{11}^{3}\left(  2l+1\right)  -(\Delta_{11}^{3})^{\ast}\left(
2l+1\right)  \right]
\end{equation}
(where $\sum_{m=-l}^{l}\sqrt{1-m^{2}/l^{2}}\simeq1+1.58l$). Thus
$\overline{I_{3}(l+1,m^{\prime};l,m)}$ contains terms that scale as $l^{-2}$
and $l^{-1}$. We then have an $m$-averaged consistency relation%
\begin{equation}
\overline{\left\langle a_{(l+1)m^{\prime}}^{\ast}a_{lm}\right\rangle }%
\simeq-\frac{i}{2\pi^{2}}\overline{I_{3}(l+1,m^{\prime};l,m)}(2\pi^{2}%
/c_{3})C_{l}\ .\label{consisreln_av}%
\end{equation}
This implies that, at high $l$, the $m$-averaged covariance matrix will
contain terms that scale as%
\begin{equation}
\overline{\left\langle a_{(l+1)m^{\prime}}^{\ast}a_{lm}\right\rangle }\sim
C_{l}/l^{2}\ ,\ \ C_{l}/l\ .\label{high_l_scaling}%
\end{equation}

The small-scale anisotropies are related to the large-scale anisotropies in
the sense that the same (unknown) coefficients $\Delta_{10}^{3}$ and
$\Delta_{11}^{3}$ appear in the small-scale $l-(l+1)$ correlations
(\ref{consisreln}) as appear in the term $i=3$ of the large-scale $l-(l+1)$
correlations (\ref{offdiag1}) (and the same is true after $m$-averaging, as in
(\ref{consisreln_av}) and (\ref{avcovmat2})). Furthermore, the small-scale
correlations (\ref{consisreln}) depend on the residue coefficient $c_{3}$,
which can be obtained by fitting the deficit function (\ref{numksi}) to the
current data for the power deficit (via the relation (\ref{C_l_corr}) between
the corrected $C_{l}$'s and $\xi(k)$).\footnote{The theoretical value of
$c_{3}$ depends on the (unknown) number $M$ of excited states in the
pre-inflationary era, but the actual value of $c_{3}$ may be deduced simply by
fitting (\ref{numksi}) to the data.}

The relationship between small-scale and large-scale anisotropies could be
tested observationally, at least in principle. If we could measure or
constrain the large-scale covariance matrix elements $\left\langle
a_{(l+1)m^{\prime}}^{\ast}a_{lm}\right\rangle $ to sufficient accuracy, then
from the expression (\ref{offdiag1}) we could deduce the values of the
coefficients $\Delta_{10}^{3}$ and $\Delta_{11}^{3}$ in the term for $i=3$
(where the function $F_{3}(l+1,l)$ is known). We could then compare with the
coefficients $\Delta_{10}^{3}$ and $\Delta_{11}^{3}$ appearing in the
expression (\ref{consisreln}) for the small-scale matrix elements
$\left\langle a_{(l+1)m^{\prime}}^{\ast}a_{lm}\right\rangle $ -- having
already found $c_{3}$ from a best-fit of $\xi(k)$ to the power deficit.
Clearly, there is a relationship between the $l-(l+1)$ correlations at large
and small scales regardless of the actual values of $\Delta_{10}^{3}$ and
$\Delta_{11}^{3}$. Similar reasoning applies to the $m$-averaged matrix
elements $\overline{\left\langle a_{(l+1)m^{\prime}}^{\ast}a_{lm}\right\rangle
}$. Whether or not this has any bearing on the reported correlations between
directional asymmetries at large and small scales remains to be seen.

\subsection{Mode alignment}

As we mentioned in Section 2.3, in their 2013 data release the Planck team
confirmed an anomalous alignment between the quadrupole ($l=2$) and octopole
($l=3$) modes \cite{PlanckXXIII-2013}. Could such alignment be explained in
terms of cosmological quantum relaxation?

The orientation of a multipole $l$ may be defined by the method of
maximisation of the `angular momentum dispersion' \cite{deOC04}. Taking
$a_{lm}(\mathbf{\hat{n}})$ to be the harmonic coefficients $a_{lm}$ expressed
in a rotated coordinate system with $z$-axis in the direction $\mathbf{\hat
{n}}$, one may consider%
\begin{equation}
\sum_{m}m^{2}\left\vert a_{lm}(\mathbf{\hat{n}})\right\vert ^{2}
\label{angmomdisp}%
\end{equation}
for arbitrary $\mathbf{\hat{n}}$ and find the axis $\mathbf{\hat{n}%
}=\mathbf{\hat{n}}_{l}$ for which (\ref{angmomdisp}) is maximised. One may
then take $\mathbf{\hat{n}}_{l}$ to define the multipole `orientation' (up to
a sign).

If the CMB is statistically isotropic, the orientations $\mathbf{\hat{n}}_{l}$
and $\mathbf{\hat{n}}_{l^{\prime}}$ of different multipoles $l$ and
$l^{\prime}$ will be statistically independent. Furthermore, each
$\mathbf{\hat{n}}_{l}$ will be a random variable drawn (independently) from a
probability distribution such that all possible directions are equally likely.
The random variable $\mathbf{\hat{n}}_{l}\cdot\mathbf{\hat{n}}_{l^{\prime}}$
will be uniformly distributed on $[-1,1]$. If we do not distinguish between
$\mathbf{\hat{n}}$ and $-\mathbf{\hat{n}}$ we may consider $\left\vert
\mathbf{\hat{n}}_{l}\cdot\mathbf{\hat{n}}_{l^{\prime}}\right\vert $, whose
probability distribution will be uniform on $[0,1]$. The data indicate that
$\mathbf{\hat{n}}_{2}$ and $\mathbf{\hat{n}}_{3}$ are improbably aligned, with
a value $\left\vert \mathbf{\hat{n}}_{2}\cdot\mathbf{\hat{n}}_{3}\right\vert
\simeq0.9849$ \cite{deOC04}. For a statistically isotropic sky, such close
alignment occurs with a probability of $1-0.9849\simeq1/66$. (Consistent
results for the alignment are obtained \cite{PlanckXXIII-2013} using an
alternative method based on the multipole vector decomposition \cite{Copi04}.)

If instead the primordial perturbations are not statistically isotropic, then
the orientations $\mathbf{\hat{n}}_{l}$ and $\mathbf{\hat{n}}_{l^{\prime}}$ of
different multipoles need not be statistically independent and all possible
directions for each $\mathbf{\hat{n}}_{l}$ need not be equally likely.
According to the scenario discussed in this paper, there will be statistical
anisotropy at small $k$ with approximate isotropy at large $k$. We are then
more likely to find anomalous statistics for $\mathbf{\hat{n}}_{l}$ at lower
values of $l$ than we are to find them at higher values of $l$.

To make this quantitative, we would have to consider primordial probability
distributions that are consistent with our anisotropic power spectrum
(\ref{ksi_aniso}) and calculate the resulting probability for multipole
alignment as a function of $l$. In particular, we would need to evaluate the
probability of finding significant alignment at the lowest $l$ values with no
significant alignment at higher $l$. This would provide a further constraint
on our model from the data.

In quantum equilibrium, of course, we expect no alignment for any $l$: the
orientations $\mathbf{\hat{n}}_{l}$ and $\mathbf{\hat{n}}_{l+1}$ will be
independent random variables distributed uniformly on the unit sphere and
$\left\vert \mathbf{\hat{n}}_{l}\cdot\mathbf{\hat{n}}_{l+1}\right\vert $ will
be distributed uniformly on $[0,1]$. In quantum nonequilibrium, however, for
low $l$ our non-vanishing matrix elements $\left\langle a_{(l+1)m^{\prime}%
}^{\ast}a_{lm}\right\rangle \neq0$ indicate that $\mathbf{\hat{n}}_{l}$ will
not be statistically independent of $\mathbf{\hat{n}}_{l+1}$ and so
$\left\vert \mathbf{\hat{n}}_{l}\cdot\mathbf{\hat{n}}_{l+1}\right\vert $ may
be distributed non-uniformly on $[0,1]$. We have seen that the averaged
correlation $\overline{\left\langle a_{(l+1)m^{\prime}}^{\ast}a_{lm}%
\right\rangle }$ decays rather rapidly as $\sim l^{-5/2}$ or $\sim l^{-3}$, so
we might reasonably expect significant alignment of $\mathbf{\hat{n}}_{l}$ and
$\mathbf{\hat{n}}_{l+1}$ only at the lowest values of $l$.

A proper assessment, however, requires us to construct appropriate primordial
distributions that are consistent with the anisotropic spectrum
(\ref{ksi_aniso}) and to evaluate the corresponding (nonequilibrium)
probability distributions for $\left\vert \mathbf{\hat{n}}_{l}\cdot
\mathbf{\hat{n}}_{l+1}\right\vert $. We leave this task for future work.

\section{Conclusion}

In this paper we have shown that cosmological quantum relaxation predicts an
anisotropic primordial power spectrum $\mathcal{P}_{\mathcal{R}}%
(k,\mathbf{\hat{k}})$ equal to the usual isotropic quantum equilibrium
spectrum $\mathcal{P}_{\mathcal{R}}^{\mathrm{QT}}(k)$ modulated by an
anisotropic deficit function $\xi(k,\mathbf{\hat{k}})$. In the limit of weak
anisotropy we have derived an explicit expression for $\xi(k,\mathbf{\hat{k}%
})$ given by (\ref{ksi_aniso}). We have explored some of the potentially
testable consequences. The function $\xi(k,\mathbf{\hat{k}})$ is determined
(up to arbitrary constants) by the functions (\ref{fns}), which are in turn
determined by derivatives of the inverse-tangent power deficit (\ref{numksi})
which was derived previously from numerical simulations. The lowest-order
off-diagonal corrections to the CMB covariance matrix are similarly determined
and at low $l$ the $l-(l+1)$ inter-multipole correlations show a
characteristic scaling with $l$ that is related to the inverse-tangent
deficit. Our anisotropic spectrum also predicts a residual statistical
anisotropy in the limit of large $l$, for which we have derived an approximate
consistency relation between the scaling of the $l-(l+1)$ correlations and the
scaling of the angular power spectrum $C_{l}$. Furthermore, the $l-(l+1)$
correlations at large and small scales are found to be related. We have
sketched how our results might be applied to understanding the anomalous
alignment of very low-$l$ multipoles, though this requires further study.

The key physical argument leading to these results was presented in Section
3.3. The quantum relaxation dynamics is independent of the direction
$\mathbf{\hat{k}}$ of the wave vector for the relaxing field mode, hence we
expect to find the same inverse-tangent deficit function (\ref{numksi}) for
different directions $\mathbf{\hat{k}}$ but in general with $\mathbf{\hat{k}}%
$-dependent fitting coefficients $c_{1}(\mathbf{\hat{k}})$, $c_{2}%
(\mathbf{\hat{k}})$ and $c_{3}(\mathbf{\hat{k}})$ (assuming there are
different initial conditions for different $\mathbf{\hat{k}}$'s) resulting in
an effective $\mathbf{\hat{k}}$-dependent or anisotropic deficit function
$\xi(k,\mathbf{\hat{k}})$. By expanding the coefficients $c_{1}$, $c_{2}$,
$c_{3}$ as functions of $\mathbf{\hat{k}}$ and assuming weak anisotropy, to
lowest order we are able to derive an explicit expression for $\xi
(k,\mathbf{\hat{k}})$ and thereby relate the off-diagonal covariance matrix to
(derivatives of) the inverse-tangent deficit function. That is the essence of
the argument made in this paper, from which the predictions follow.

We have not considered how realistic it might be to test our predictions. At
low $l$ we predict certain scalings $\sim l^{-5/2},\ l^{-3}$ for the
$m$-averaged covariance matrix elements $\overline{\left\langle
a_{(l+1)m^{\prime}}^{\ast}a_{lm}\right\rangle }$, while at high $l$ we predict
an approximate consistency relation $\overline{\left\langle a_{(l+1)m^{\prime
}}^{\ast}a_{lm}\right\rangle }\sim C_{l}/l^{2},\ C_{l}/l$ between the scaling
of $\overline{\left\langle a_{(l+1)m^{\prime}}^{\ast}a_{lm}\right\rangle }$
and the scaling of the angular power spectrum $C_{l}$. Are we likely to find
statistically-significant evidence for or against such scalings in the CMB
data? The evidence for $l-(l+1)$ correlations themselves seems fairly
significant, but we require statistical tests searching for a particular
scaling of the correlations with $l$. Perhaps appropriate tests could be
devised. Any resulting significance may be slight when considered alone, but
the overall significance could be raised when combined with other related
features such as the power deficit and the small-scale anisotropies. It does
not seem out of the question that, by combining these features and considering
the predicted relationships between them, the overall statistical significance
for or against cosmological quantum relaxation could turn out to be
considerable. However, we leave a proper analysis for future work.

In our scenario there is a relationship between the $l-(l+1)$ correlations at
large and small scales because the same unknown coefficients $\Delta_{10}^{3}$
and $\Delta_{11}^{3}$ appear in the large-scale covariance matrix elements
(\ref{offdiag1}) and in the small-scale covariance matrix elements
(\ref{consisreln}). In future work it would be of interest to examine if the
predicted relationship could play a role in explaining the reported
correlations between directional asymmetries at large and small scales
\cite{Planck15-XVI-IsoStats}.

It is instructive to compare our results with some of the other theoretical
work on statistical anisotropy in the CMB.

A systematic and model-independent discussion of anisotropic primordial power
spectra of the general form (\ref{anisoPS}) was presented by Armendariz-Picon
\cite{AP06}, who considered in particular coefficients $g_{LM}(k)$ that are
power laws in $k$. Armendariz-Picon also considered anisotropies with a
cut-off at some maximum multipole and for this case showed how to construct an
unbiased estimator that could be used to measure the values of certain
integrals which are related to our integrals (\ref{F_i}). Ackerman, Carroll
and Wise \cite{ACW07} considered imprints of a primordial preferred direction
on the CMB, assuming a parity $\mathbf{k}\rightarrow-\mathbf{k}$ symmetry so
that the leading term in (\ref{anisoPS}) has $L=2$ (yielding off-diagonal
terms $\left\langle a_{(l\pm2)m^{\prime}}^{\ast}a_{lm}\right\rangle \neq0$ in
the covariance matrix), and furthermore they presented a model in which the
corresponding coefficients are scale-invariant (independent of $k$) in the
region of interest. Similarly, Pullen and Kamionkowski \cite{PK07} discussed
CMB statistics for an anisotropic primordial spectrum in which the general
form (\ref{anisoPS}) is restricted to terms with $L$ even (so that in
particular there is no dipolar term with $L=1$). Ma, Efstathiou and Challinor
\cite{MEC11} discussed the effects of a quadrupole ($L=2$) modulation of the
primordial spectrum, with a power-law dependence on $k$, again assuming a
parity-invariance to rule out odd values of $L$. A notable recurring feature
of the cited work is the assumption that the lowest-order $L=1$ term (and
indeed all odd-$L$ terms) may be discarded on grounds of parity symmetry. As a
result, these models do not yield $l-(l+1)$ correlations in the CMB (for which
there is some evidence). In contrast, when considering the possibility of
primordial quantum nonequilibrium, we should bear in mind that the resulting
departures from the Born rule can in principle be quite arbitrary. In
pilot-wave dynamics, the initial probability distribution is a contingent fact
about the world to be determined empirically; it is in principle unconstrained
by physical law. Therefore, there are no physical grounds for imposing parity
symmetry on primordial quantum nonequilibrium. It is then quite conceivable
that terms with $L=1$ will be present and could even be dominant (as we have
assumed for definiteness and simplicity). Furthermore, because of the
intrinsic contingency of nonequilibrium, we would not expect it to manifest
simply as an effective preferred direction but as a more complicated and
generalised asymmetry in $\mathbf{k}$-space.

More recently, predictive models have been proposed that break parity
symmetry. For example, refs. \cite{JAFSW14} and \cite{MS15} find dipolar terms
generated by primordial domain walls and by a direction-dependent inflaton
field respectively. We note that the various available models are based on
quite different physical mechanisms and make different predictions. For
example, ref. \cite{BPM15} predicts both a power deficit and a statistical
anisotropy arising from a pre-inflationary Kasner geometry, where at small
scales the predicted spectrum converges to the standard isotropic form. This
is in contrast with the scenario of cosmological quantum relaxation discussed
here, for which we expect a residual statistical anisotropy even at small
scales. Further comparison among the proposed models and with data is,
however, a matter for future work.

Finally, it is worth noting that definitive observational evidence for
statistical anisotropy in the primordial spectrum would not necessarily
require us to abandon the inflationary paradigm for the origin of cosmic
structure. As we have discussed, in the pilot-wave formulation of quantum
theory the probability distribution is in principle independent of the quantum
state. In particular, quantum nonequilibrium changes the spectrum of vacuum
fluctuations without changing the vacuum wave functional itself. It is then
straightforward to introduce primordial anisotropies without altering the
essential structure and formalism of inflationary cosmology. Quantum
relaxation ensures that the standard predictions are still obtained to a first
approximation. To a second approximation, the suppression of quantum
relaxation at large scales offers a single physical mechanism that might
explain both the observed power deficit and the observed statistical
anisotropy. We have shown that the dynamics of quantum relaxation leads to
quantitative predictions for both kinds of anomaly as well as predicting
relations between them. Whether these predictions will be confirmed or ruled
out by the data remains to be seen.

\textbf{Acknowledgement}. I am grateful to participants at the conference `The
Primordial Universe after Planck', held at the Institut d'Astrophysique de
Paris in December 2014, for helpful comments and discussions.

\end{document}